\documentclass[twocolumn,showpacs,preprintnumbers,amsmath,amssymb]{revtex4}

\usepackage{graphicx}

\usepackage{epsfig}

\usepackage{amssymb}
\usepackage{amsmath}

\def\url#1{{\ttfamily\def\/{/\discretionary{}{}{}}#1}}
\def\bibcode#1{}

\newcommand{\aap}{{Astron.~Astrophys.}}

\newcommand{\mnras}{{Mon.~Not.~R.~Astron.~Soc.}}

\def\bt{\begin{tabular}{@{} c @{}}}
\def\et{\end{tabular}}
\def\cl{\centerline}

\begin{document}

\title{Search for Gravitational Waves in the CMB After WMAP3:
  Foreground Confusion and The Optimal Frequency Coverage for
  Foreground Minimization} \author{Alexandre Amblard, Asantha Cooray,
  Manoj Kaplinghat}, \address{Center for Cosmology, Department of
  Physics and Astronomy, University of California, Irvine, CA 92697}

\begin{abstract}
B-modes of Cosmic Microwave Background (CMB) polarization can be
created by a primordial gravitational wave background. If this
background was created by Inflation, then the amplitude of the
polarization signal is proportional the energy density of the universe
during inflation. The primordial signal will be contaminated by
polarized foregrounds including dust and synchrotron emission within
the galaxy. In light of the WMAP polarization maps, we consider the
ability of several hypothetical CMB polarization experiments to
separate primordial CMB B-mode signal from galactic foregrounds. We
also study the optimization of a CMB experiment with a fixed number of
detectors in the focal plane to determine how the detectors should be
distributed in different frequency bands to minimize foreground
confusion. We show that the optimal configuration requires
observations in at least 5 channels spread over the frequency range
between 30 GHz and 500 GHz with substantial coverage around 150
GHz. If a low-resolution space experiment using 1000 detectors to
reach a noise level of about 1000 nK$^2$ concentrates on roughly 66\%
of the sky with the least foreground contamination the minimum
detectable level of the tensor-to-scalar ratio would be about 0.002
at the 99\% confidence level for an optical depth of 0.1.

\end{abstract}


\maketitle

\section{Introduction}

The cosmic microwave background (CMB) is a well known probe of the
early universe. The acoustic peaks in the angular power spectrum of
CMB anisotropies capture the physics of the primordial photon-baryon
fluid undergoing oscillations in the potential wells of dark matter
\cite{Huetal97}. The associated physics involving the evolution of
a single photon-baryon fluid under Compton scattering and gravity is
both simple and linear \cite{SacWol67,Sil68,PeeYu70,SunZel70}. 

The acoustic peak structure has been well established with the
Wilkinson Microwave Anisotropy Probe (WMAP;
\cite{Hinetal06,Speetal06}) data.  Over the next several years, with
the launch of the ESA's Planck surveyor, it is expected that the
anisotropy power spectrum will be measured to the cosmic variance
limit out to an angular scale of roughly ten arcminutes (or multipoles
$\sim$ 2000 in the angular power spectrum; see, review in
\cite{HuDod02}).

Beyond temperature anisotropies, the focus is now on the polarization
at medium to larger angular scales.  When the Universe was reionized,
the temperature quadrupole rescattered at the reionization surface 
producing a new contribution to the polarization \cite{Zal97}.  Such a
signature has now been detected in the WMAP data
\cite{Pagetal06,Speetal06}. 
Furthermore, the CMB polarization field contains a signature of
primordial gravitational waves, which is considered a smoking-gun
signature for inflation as models of inflation predict a stochastic
gravitational wave background in addition to the density perturbation
spectrum \cite{Starobinsky79,KamKos99}.  The distinct signature is in the form of a
curl component, also called B-mode, of the two-dimensional
polarization field \cite{Kametal97,SelZal97}.

Observation of the primordial B-modes is now the primary goal of a
large number of ground- and balloon-borne CMB polarization
experiments. There are also plans for a next generation CMB
polarization mission as part of NASA's {\it Beyond Einstein} 
program ({\it Inflation Probe}). The only known source for {\it
primordial} B-modes is inflationary gravitational waves (IGWs) and a
detection of this tensor component would be one of the biggest
discoveries in science. The tensor component captures important
physics related to the inflaton potential, especially the energy scale
at which relevant models exit the horizon \cite{KamKos99}.

CMB polarization observations are significantly impacted by
foreground polarized radiation, with initial estimates suggesting that
the minimum amplitude to which a gravitational wave background can be
searched is not significantly below a tensor-to-scalar ratio of
10$^{-3}$ \cite{Verde06,Amarie:2005in,Carretti:2006us}.  This
limit based on analysis of foregrounds is well above the ultimate
limit of a tensor-to-scalar ratio around $10^{-4}$ due to the cosmic
shear confusion, associated with secondary 
B-modes generated from E-modes by lensing due to intervening
structure \cite{Kesden,Song,Seljak:2003pn}. If the optical depth is
around 0.1, then the ultimate limit may be pushed down an order 
of magnitude \cite{KapKnoSon03} by measuring the large angle B mode
polarization signal. 

While the impact of foregrounds is now well appreciated, it is not
fully clear how these foregrounds may be minimized in the next
generation experiments. This issue critically impacts the planning of 
experiments. For the {\it Inflation Probe} or a ground-based
experiment that attempts to target the gravitational wave background, 
one of the most significant issues to address is the choice of
frequency bands for observations such that the foreground
contamination is minimized. For example, should an experiment target
the low-frequency end where synchrotron dominates, high-frequency end
where dust dominates, or the frequency range between 50 GHz to 100 GHz
where foreground polarization is minimum, but contributions from both
synchrotron and dust are expected? 
We find that frequency bands over a wide range from low frequencies
dominated by synchrotron to high frequencies dominated by dust are
required.  Furthermore, we also study the optimization problem of
dividing a fixed number of detectors in the focal plane among the
frequency bands. We looked into the question of whether the division
should be such that one has equal noise in each band or whether we
should concentrate more detectors in channels where foreground
components dominate. 

To address the issue of foreground contamination and the optimization
necessary to minimize foregrounds, we made use of the recently 
released WMAP polarization data \cite{Pagetal06}. The WMAP data has
provided us with an estimate of the polarized synchrotron foregrounds
at low frequencies. For dust polarization, we also made use of dust
maps from Ref.~\cite{SFD98}.  We considered several different
experimental possibilities with frequency coverages either at the
high-end or low-end of our frequency range as well as an experiment
with several channels that cover the range from 30 GHz to 300
GHz. 

The paper is organized in the following manner. In the next section,
we will discuss the simulated maps, summarize the foreground model and
discuss the foreground removal method employed. In Section~3, we will
discuss contamination in example CMB experiments with two or
more frequency bands spread over the broad range from 30 GHz to 300 
GHz. In Section~4, we will consider the optimization of 
experiments such as {\it Inflation Probe}. Here optimization refers to
the selection of frequency bands with total number of detectors kept
fixed so as to minimize foreground confusion. We
discuss our results and conclude with a summary in Section~5. 

Our main results are two-fold. First, we quantify the effect
of foregrounds on experiments with different frequency coverages and the
improved sensitivity to B modes from adding future WMAP or Planck
data. Our results from this study are summarized in Table
\ref{tab:rmin}. 
The second result concerns the optimal spacing of frequency bands for a
CMB experiment that aims to detect primordial B-modes. We discuss the
optimization in \S~III while our results are summarized in Table
\ref{tab:freqopti}. 

\begin{table}[!ht]
\begin{center}
{\scriptsize
\begin{tabular}{cccccc}
\hline
Expt.$^1$ & Frequencies & NET$^2$  & \bt angular\\ resolution\et &
f$_\mathrm{sky}$ & T$_{\rm obs}$ \\
& (GHz) &  ($\mu$K$\sqrt{\rm sec}$) &  (arcmins) & & (days) \\
\hline \hline
A & \bt 45, 75, 85,\\[-3pt] 100, 145, 165 \et & \bt 13.0, 6.0, 5.6,\\[-3pt]
6.2, 5.0, 5.5 \et & \bt 115.2, 69.1, 60.4,\\[-3pt] 52.4, 36.0, 32.0
\et & \bt 45\% \\ [-3pt] (34\%) \et & 10\\\hline
B & 100, 150, 220 & 9.8, 10.4, 35.2 & 55, 37, 26  & 2.5\% & 600\\\hline
C & 40, 90 & 18, 9 & 16, 7 & 3\% & 1000 \\\hline
D & \bt 40, 60, 90 \\[-3pt] 135, 200, 300 \et & 
\bt 12.0, 7.1, 3.4 \\[-3pt] 2.7, 2.7, 3.4 \et & \bt 116,
77.5, 52.0 \\[-3pt] 34.5, 23.0, 15.5 \et & \bt 100\% \\ [-3pt] (13\%) \et & 730\\\hline
\end{tabular}}
\caption{Example CMB polarization experiments used for the
foreground analysis and their specifications in terms of the
experimental noise, angular resolution and the sky area observed. \\
Notes: ---\\ $^1$: Experiments B and C are typical of ground-based
experiments that target a small area on the sky with limited
frequencies and concentrating on either low- or high-frequencies.
Experiment A is an example of a balloon-borne experiment with a large
sky coverage. Note that the sky area observed is 45\% but we assume
that a smaller area (34\%) is clean enough for CMB measurements.
Experiment D is consistent (same as A, 100\% observed but 13\% used)
with one of the concept study designs of the {\it Inflation Probe}
mission for high precision CMB polarization measurements.\\ $^2$: The
NET (noise-equivalent-temperature) in units of $\mu$K$\sqrt{\rm sec}$
is the focal plane sensitivity at each of the channels. This is
equivalent to the NET of a single detector divided by the square-root
of the number of detectors.}
\label{tab:exppa}
\end{center}
\end{table}

\begin{figure}[!h]
\centerline{
\psfig{file=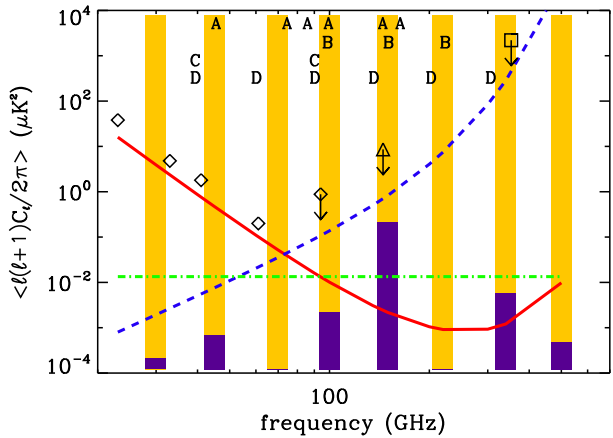,width=9cm}
}
\caption{Dust (blue dashed line), Synchrotron (red solid line) and
CMB (green dotted-dashed line) spectrum estimated with our model in
thermodynamic units (in $\mu$K$^2$). The ordinate is the average of
C$_\ell$($\ell$+1)$\ell$/2$\pi$ over $10<\ell<100$. The channels
marked with letters A, B, C, D show different experiments studied.
These channel selections represent some of the choices considered by
ongoing and planned experiments in the field. Using height as a
representation of the number of detectors, the lower bars show the
optimal experiment that minimizes foregrounds and maximizes the
detectability of primordial tensors in the B-mode polarization power
spectrum. The optimization involved a priori selection of 8 channels
shown here and a fixed number of detectors in the focal plane.  The
150 GHz channel with the largest number of detectors provide high
sensitive CMB observations, while the low- and high-frequency bands
monitor the synchrotron and dust foregrounds, respectively. The
achievable tensor-to-scalar ratio in the optimized setup is on average 36\%
better without lensing and 20\% better with lensing, than the
case where all detectors are spread equally among the 8 channels. The
foreground residual is in both cases smaller by roughly a
factor of two.  The diamond, triangle and square represent the B-mode
measurement or 2-$\sigma$ upper limit (with the down arrow) of WMAP
(value at $\ell=5$ \cite{Pagetal06}), Boomerang (value between $\ell$
of 201 and 1000, \cite{Monetal06}) and Archeops (value between $\ell$
of 20 and 70 \cite{Ponetal05}). The comparison between these
measurements and our spectra is to be treated with caution since the
scales ($\ell$ range) and the sky coverage do not always match.}
\label{fig:spectra}
\end{figure}

\begin{figure*}[!t]
\centerline{
\psfig{file=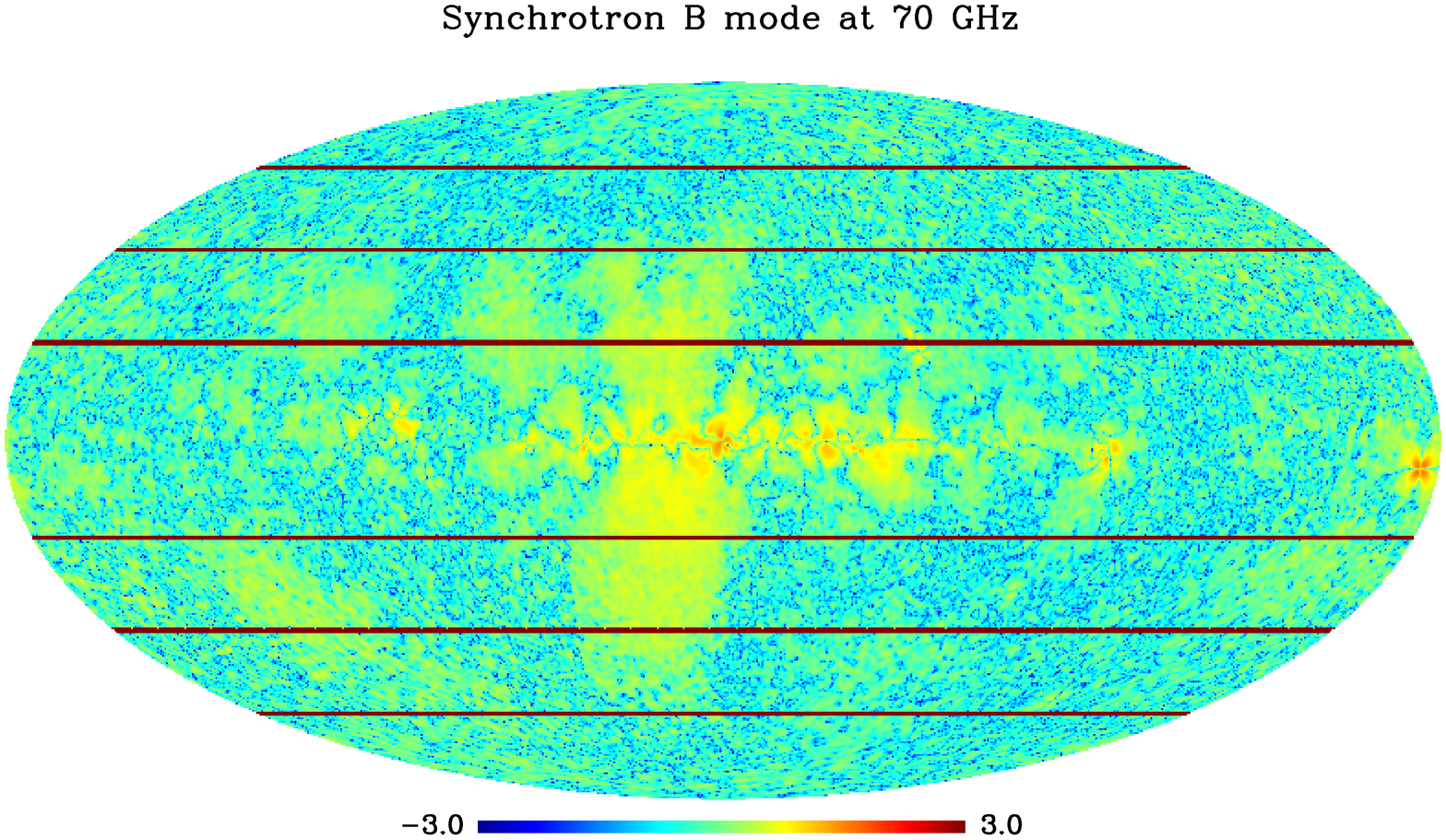,width=4.5cm,angle=90}
\psfig{file=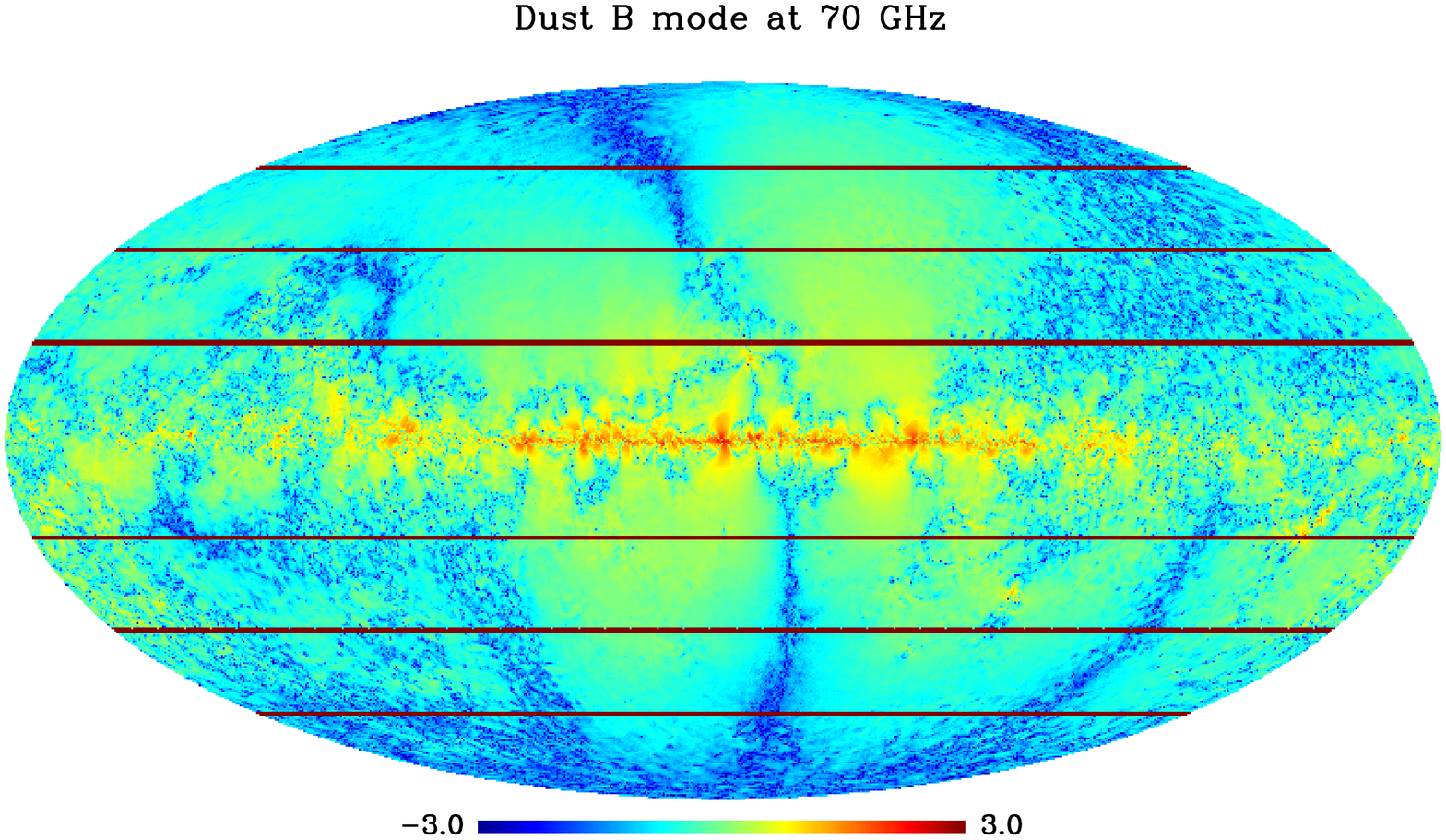,width=4.5cm,angle=90}
}
\caption{B-mode maps of the synchrotron (right) and dust (left)
emission at 70 GHz, the scale is logarithmic going from 10$^{-3}$
$\mu$K to 10$^3$ $\mu$K. The 6 horizontal lines represent the galactic
latitude at -60, -40, -20, 20, 40 and 60 degrees. We used only the part
of the map outside the galactic plane ($|$ galactic latitude$|>$20),
which represents 65.8\% of the sky. For some of the experiments we
discuss here, we also considered smaller sky coverages from 2.5\% 
to 45\%. In these cases, we assumed that the experiments will target a
low, but not necessarily the lowest, foreground emission sky area for
observations.}
\label{fig:syncdustmaps}
\end{figure*}

\section{Impact of Foregrounds on B-mode Measurements}

\subsection{Models of dust and synchrotron polarization}

We have included the new polarized data from WMAP at 23 GHz
\cite{Pagetal06,Hinetal06} in our simulations.  These observations
form the core of our calculations. We assume the WMAP 23 GHz channel
is dominated by synchrotron emission.  To avoid excess noise at
smallest angular scales probed by WMAP at 23 GHz, we filter the maps
at multipoles more than 40 and extrapolate the synchrotron power
spectrum out to a multipole of 200 using the same power-law slope for
synchrotron power spectrum with multipole at $\ell < 40$.  For
synchrotron emission at higher frequencies, we extrapolated this map
using the software provided by the WOMBAT project
\footnote{http://astro.berkeley.edu/dust/cmb/data/data.html}. The
WOMBAT project uses the spectral index $\beta$ obtained from combining 
the Rhodes/HartRAO 2326 MHz survey \cite{Jonetal98}, the Stockert 21cm
radio continuum survey at 1420 MHz \cite{Rei82,ReiRei86}, and the
all-sky 408 MHz survey \cite{Hasetal81}. 

In our extrapolation, we assumed that the spectral index varies across
the sky (down to about 1 degree) but is constant over the range of
frequencies considered.  There is, however, an indication in the WMAP
data that the synchrotron spectral index is decreasing with increasing
frequency
\cite{Hinetal06,Pagetal06}, but it is not yet established whether this
is real or a reflection of the increasing dust contribution at higher
frequencies. 

In order to simulate the dust polarization, we again made use of the
WMAP 23 GHz map by assuming that the synchrotron signal is a good
tracer of the galactic magnetic field and that the dust grains align
very efficiently with this magnetic field.  We used the synchrotron
polarization angle to describe the dust polarization as well,
consistent with the model presented by Ref.~\cite{Pagetal06} to
describe the galactic synchrotron map. For the intensity, we crudely
assumed a constant overall polarization fraction of 5\% relative to
the total dust intensity at a given frequency. Such a fraction is
consistent with most recent measurements of dust polarization outside
the galactic plane such as {\it Archeops} experiment at 353 GHz
\cite{Ponetal05}, and with theoretical models \cite{Mar06}. Using this
fraction, we again used the interpolation of model 8 of
Ref.~\cite{Finetal99} of the maps from Ref.~\cite{SFD98} to simulate
the polarized dust emission over the frequency range of 30 GHz to 300
GHz.

In the present work, we consider polarized foregrounds due dust and
synchrotron emission. Free-free  emission contributes to intensity
anisotropies but is unpolarized.  A third component in polarization
maps may come from the spinning dust. Relative to the total intensity
of the spinning dust background, the polarized fraction is about 5\%
at a few GHz but below 0.5\% at frequencies around 30 GHz, where CMB
observations begin \cite{LazDra00}.  There is no evidence for a
spinning dust component in WMAP polarization data \cite{Pagetal06},
but this is perhaps susceptible to model uncertainties.

\begin{figure}[!h]
\centerline{
\psfig{file=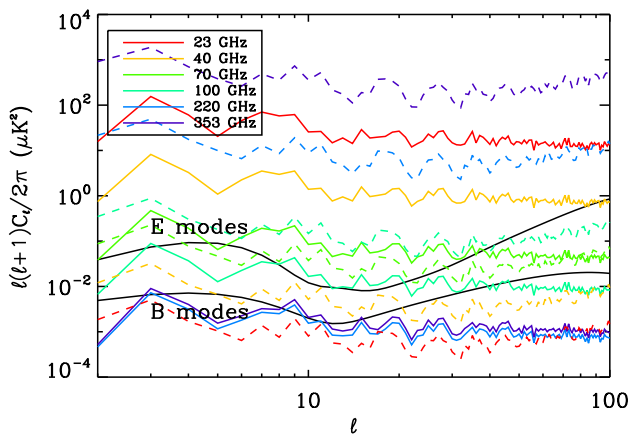,width=9cm}
}
\caption{Dust (dashed lines) and synchrotron (solid lines) power 
spectra at six frequencies between 23 and 353 GHz (from red to purple)
as labeled on the figure.  For comparison, we plotted the primordial
E-mode and B-mode CMB power spectra consistent with WMAP3 and assuming
$r=0.3$.}
\label{fig:syncdustpswfreq}
\end{figure}

Following the above guidelines, we simulated the synchrotron, dust and
CMB signals. We assumed a standard $\Lambda$CDM cosmological model in
agreement with WMAP recent results \cite{Speetal06}. The simulated
synchrotron and dust map were produced in the
HEALPix\footnote{http://healpix.jpl.nasa.gov/} pixelization
scheme. When plotting our results, for comparison, we also plot the
power spectrum of the B-mode tensor component assuming a
tensor-to-scalar ratio $r$ of 0.3. When we present our results,
however, we will discuss the minimum amplitude of the tensor component
that experiments will be able to detect given the presence of residual
foregrounds.  We plot example frequency maps as well as spectra in
Figs.~\ref{fig:syncdustmaps} \& \ref{fig:syncdustpswfreq}.

Our maps generate power spectra that match the values obtained from
WMAP data \cite{Pagetal06} for the synchrotron polarization with rms 
fluctuations at the level of $\simeq$ 25, 5, 2, 1 $\mu$K$^2$ at 23,
33, 40, 61 GHz between $\ell$ of 2 and 100. In terms of dust, our maps
and the resulting power spectra are again consistent with CMB
measurements at high frequencies: {\it Boomerang} data provide a limit
of 8.6 $\mu$K$^2$ at 145 GHz for average fluctuations at $201<\ell<
1000$ \cite{Monetal06} while {\it Archeops} data \cite{Ponetal05} lead
to a limit of 2200 $\mu$K$^2$ at 353 GHz over $20 <\ell< 70$ (both at
2 $\sigma$ confidence level). These limits are consistent with the
maximal level of dust in areas corresponding to these observations.

\begin{figure*}[t]
\centerline{\psfig{file=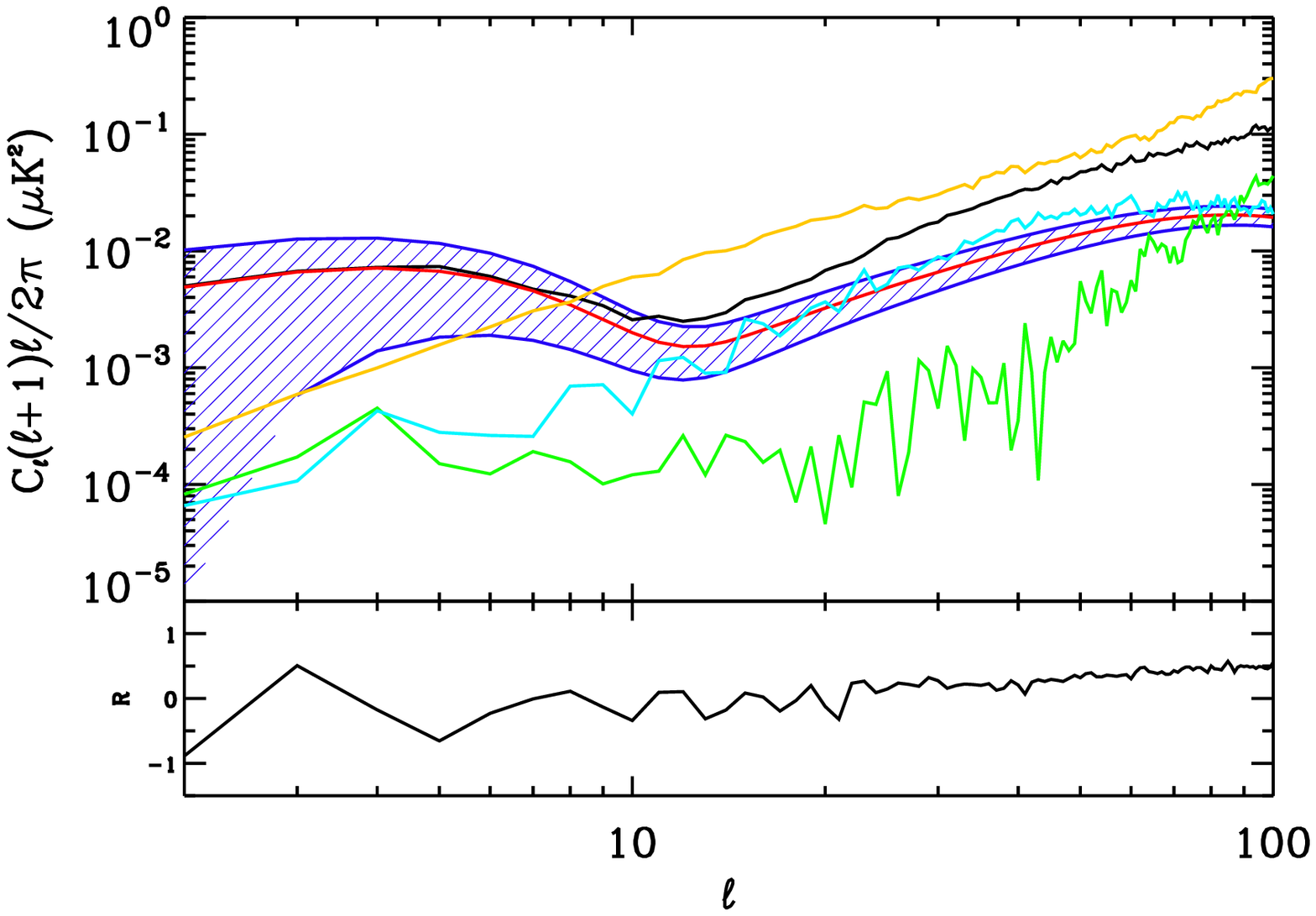,width=5cm}
\psfig{file=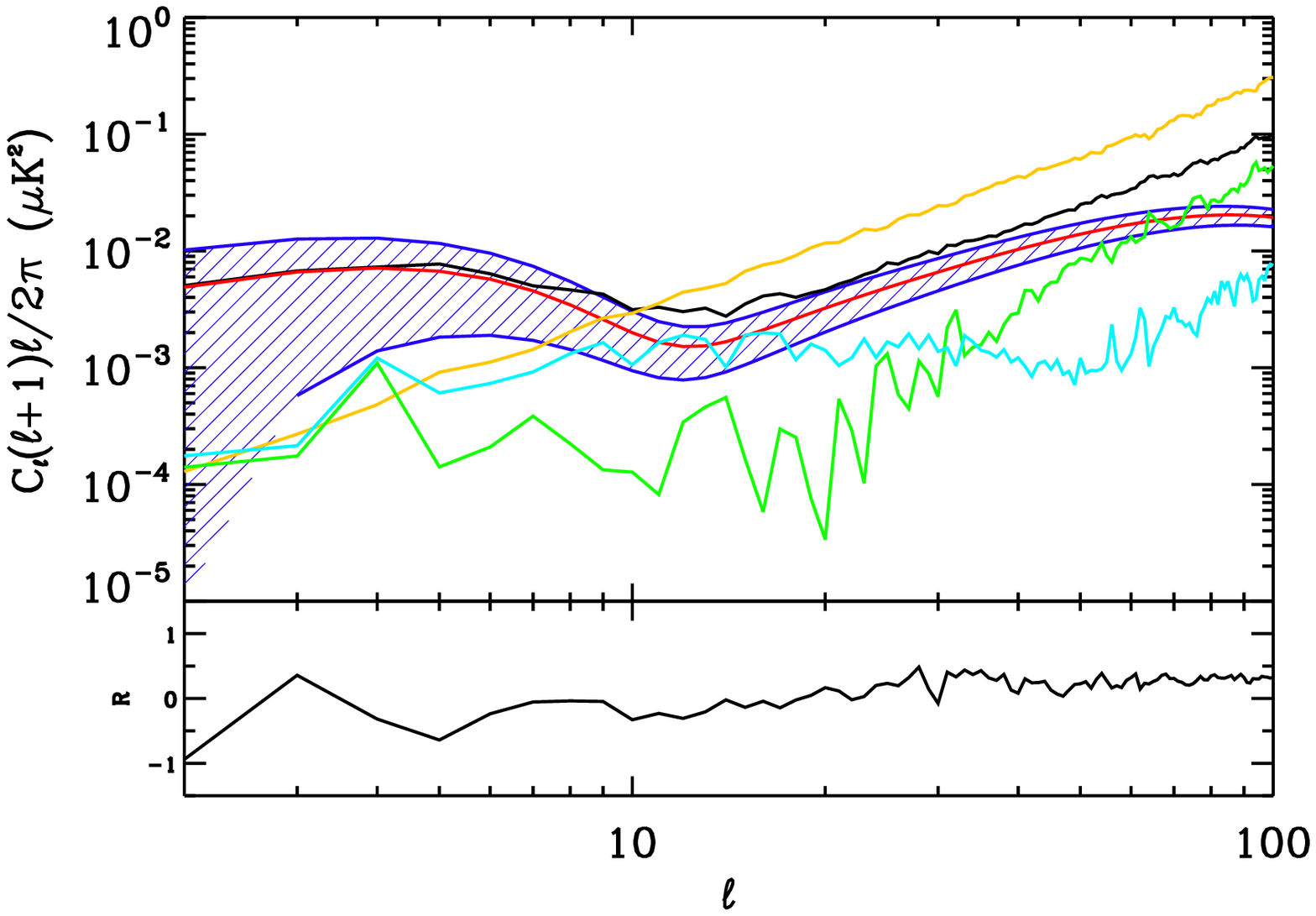,width=5cm}
\psfig{file=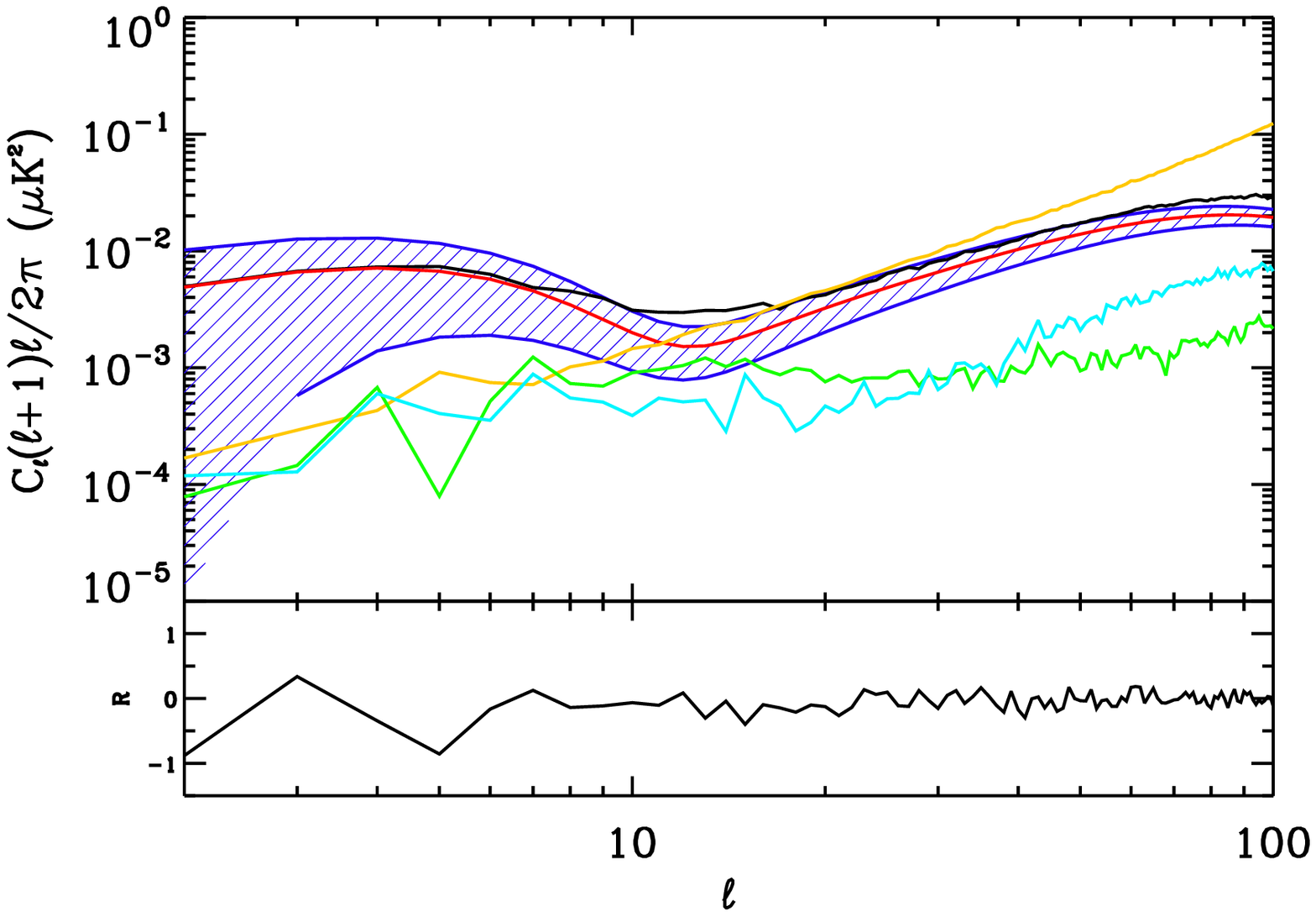,width=5cm}}
\centerline{\psfig{file=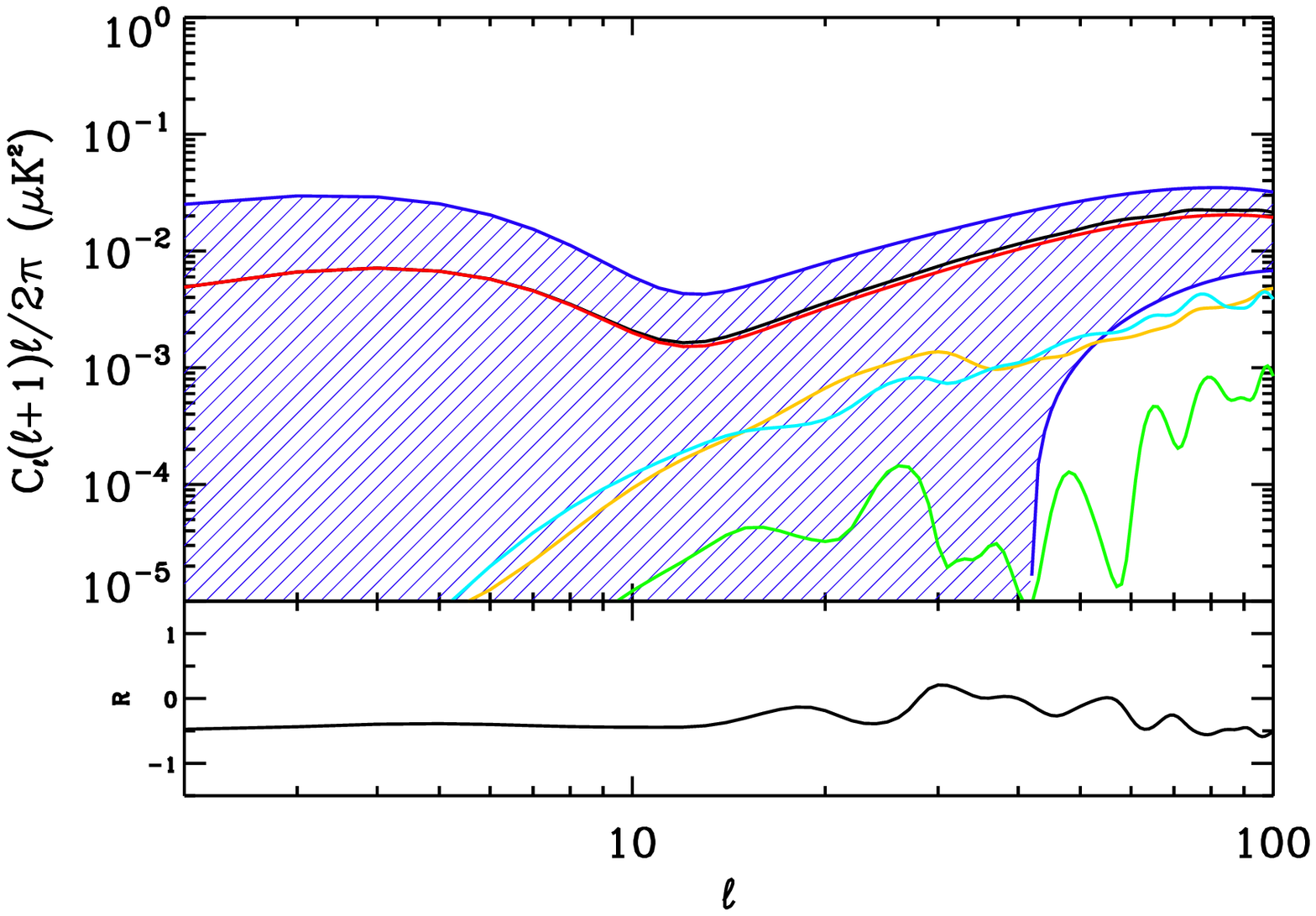,width=5cm}
\psfig{file=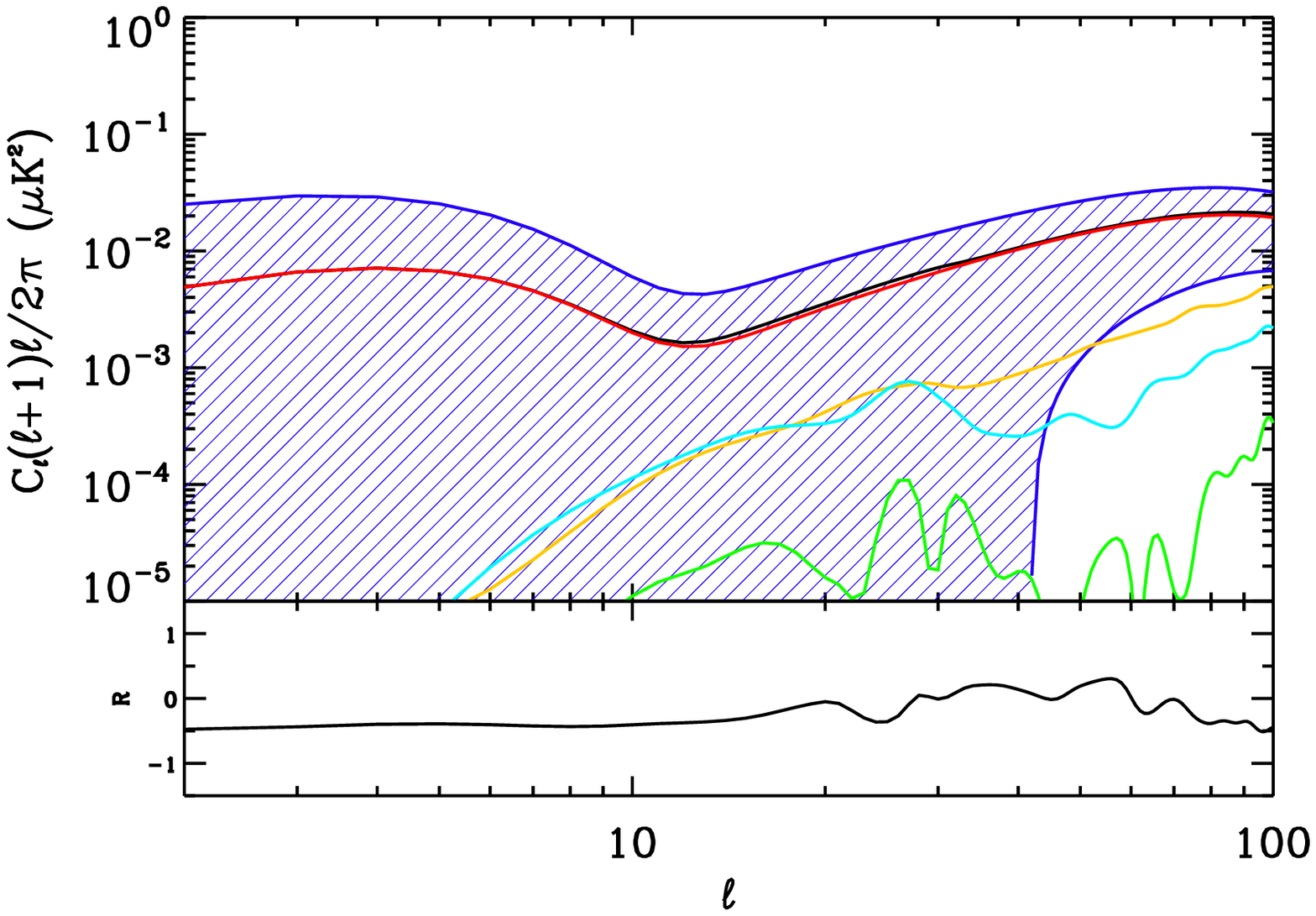,width=5cm}
\psfig{file=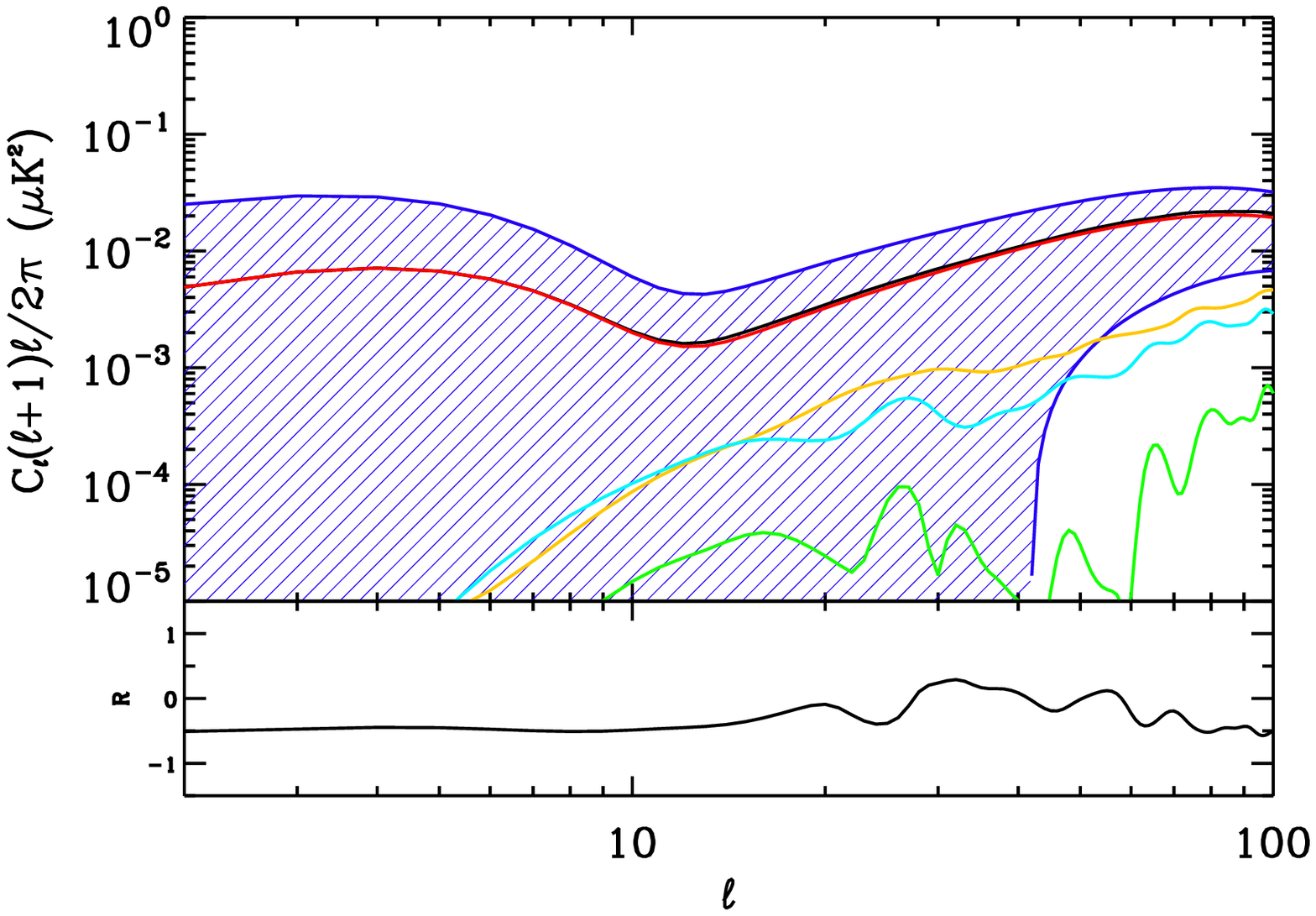,width=5cm}}
\centerline{\psfig{file=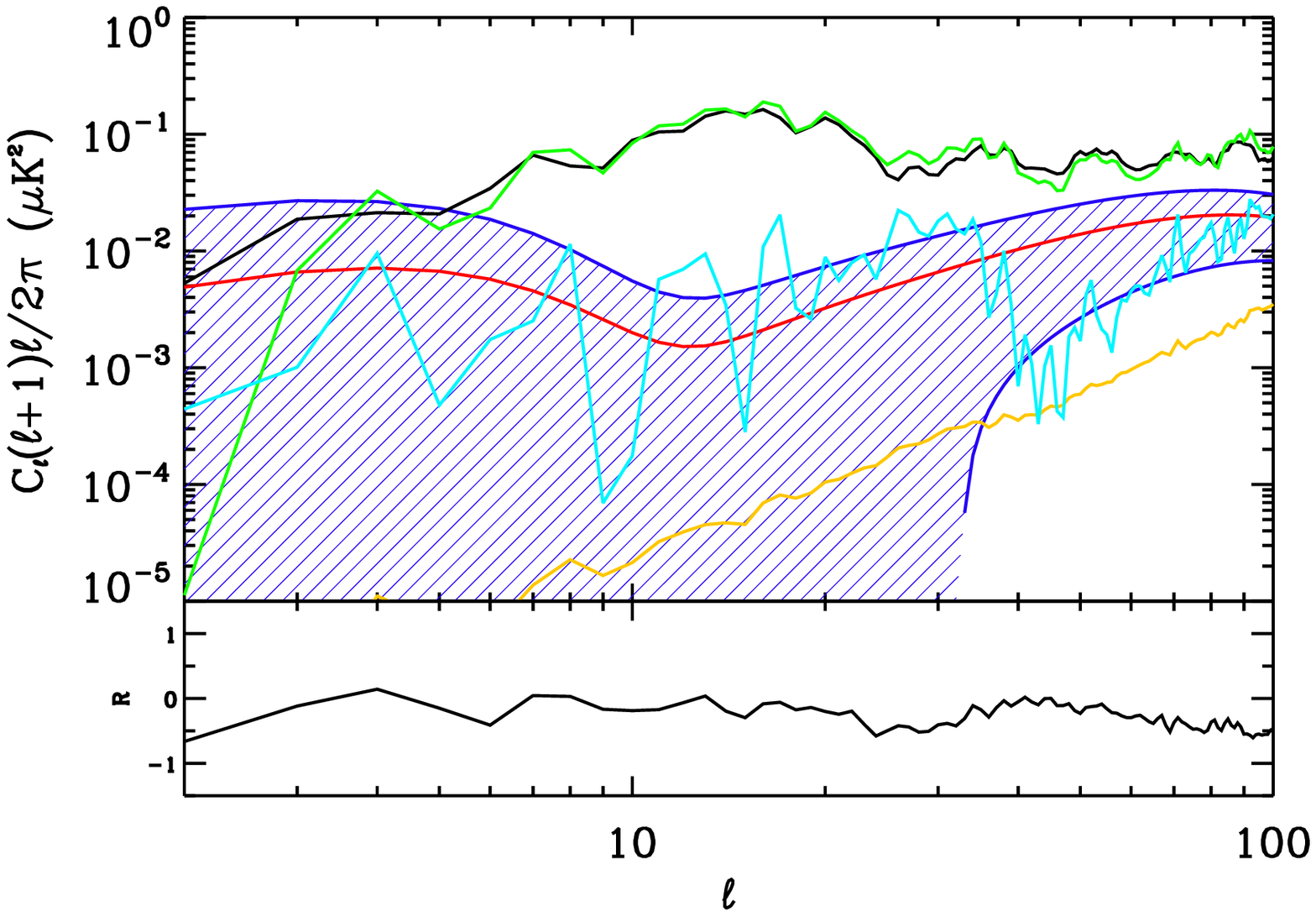,width=5cm}
\psfig{file=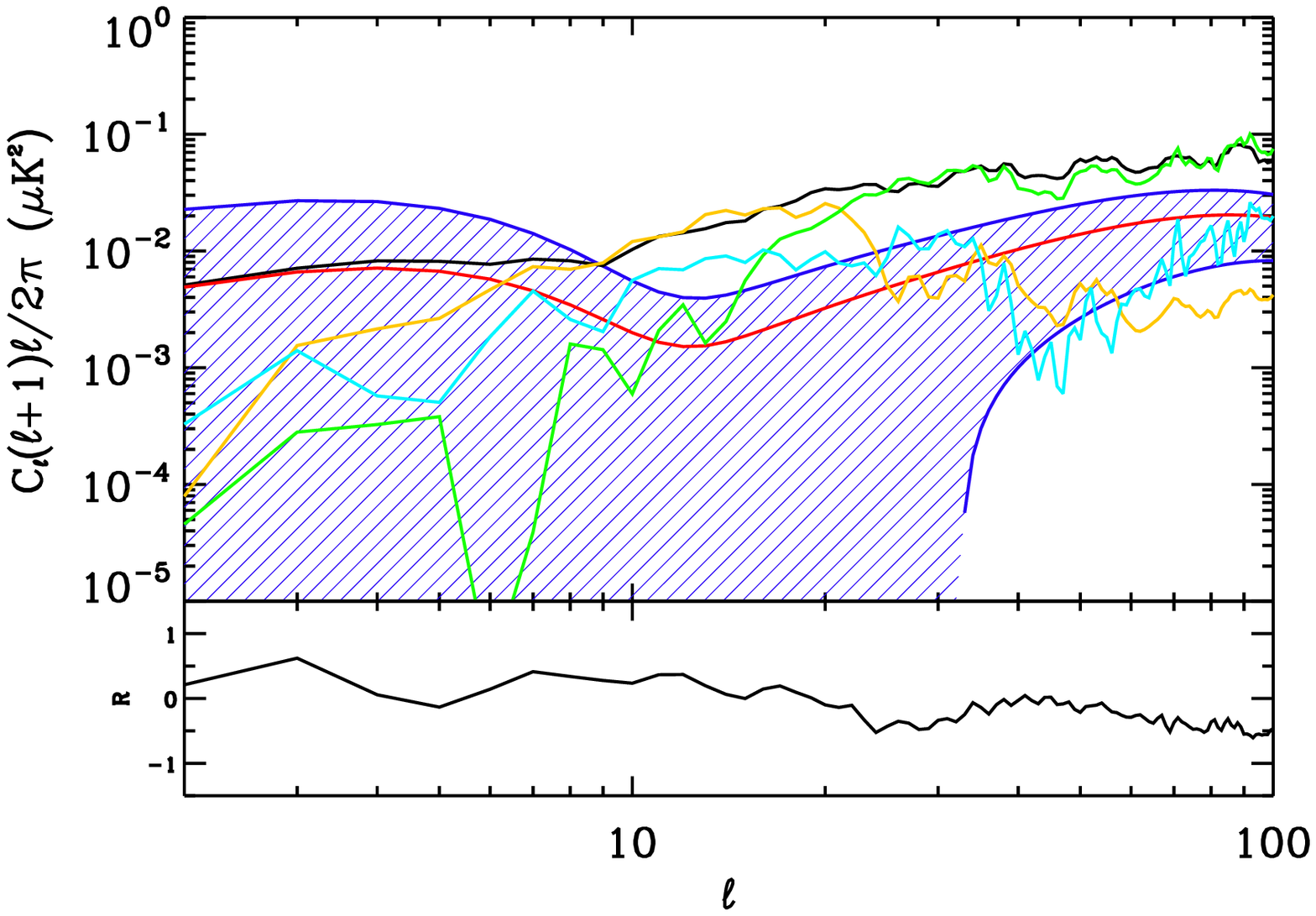,width=5cm}
\psfig{file=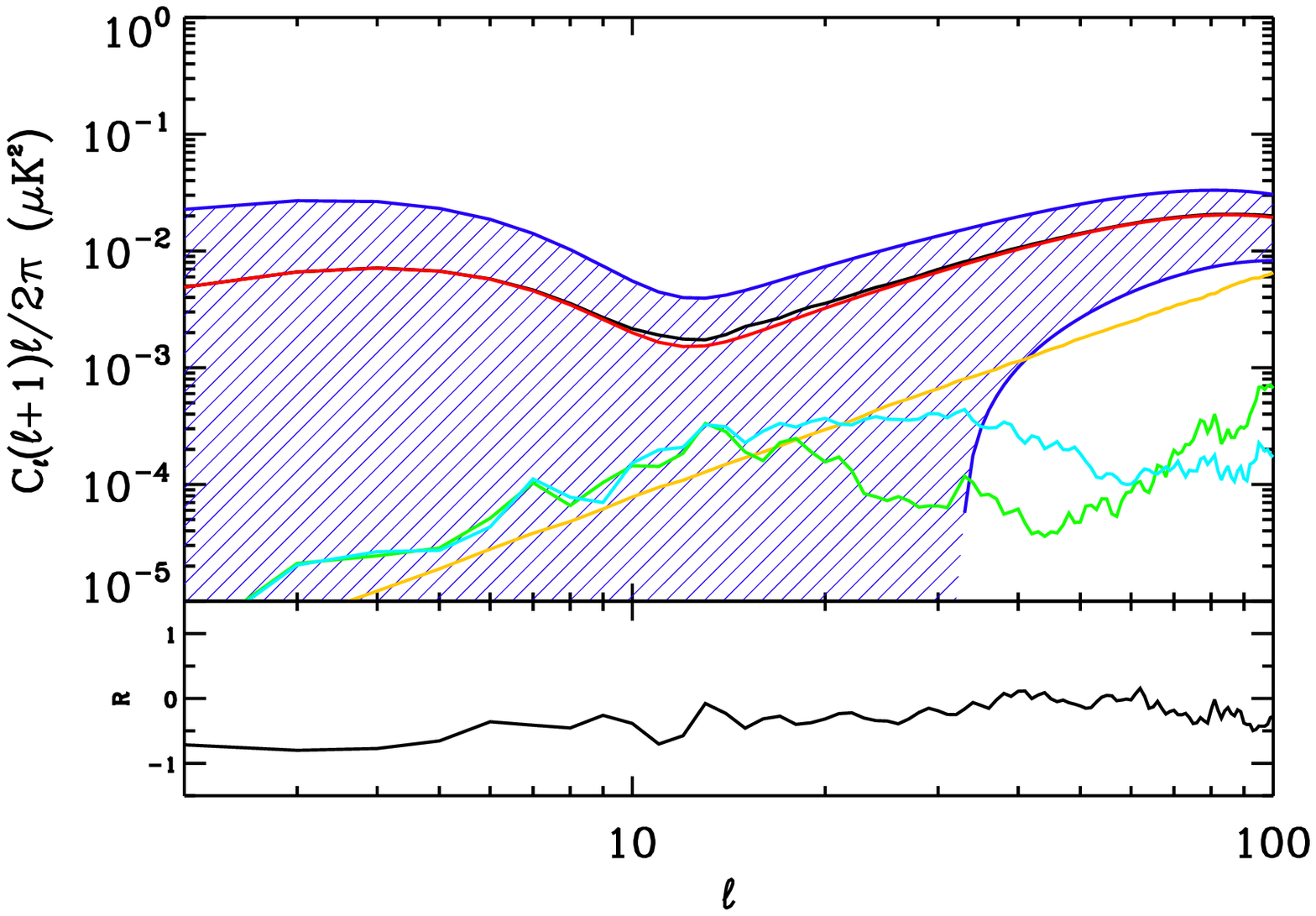,width=5cm}}
\centerline{
\psfig{file=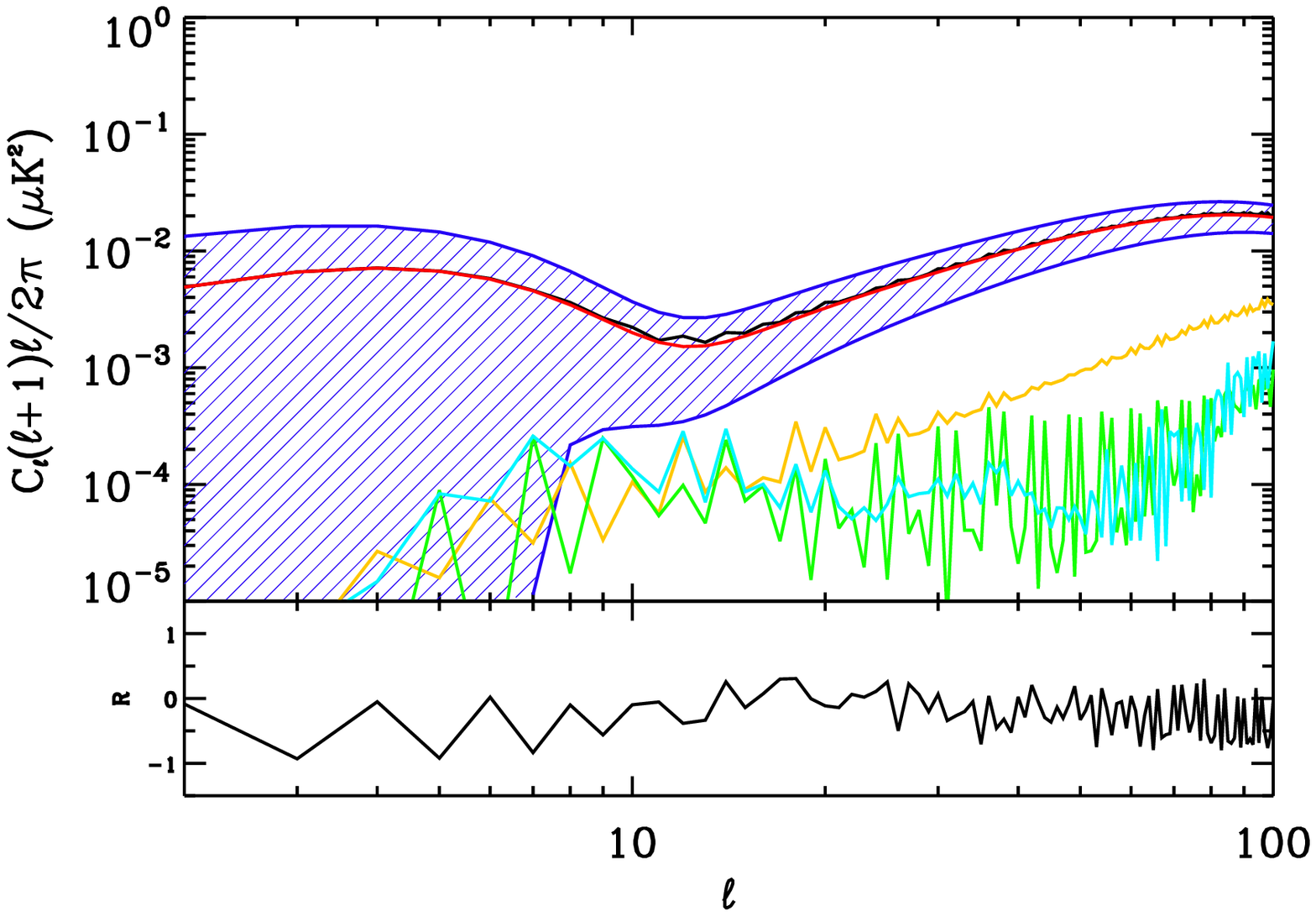,width=5cm}
\psfig{file=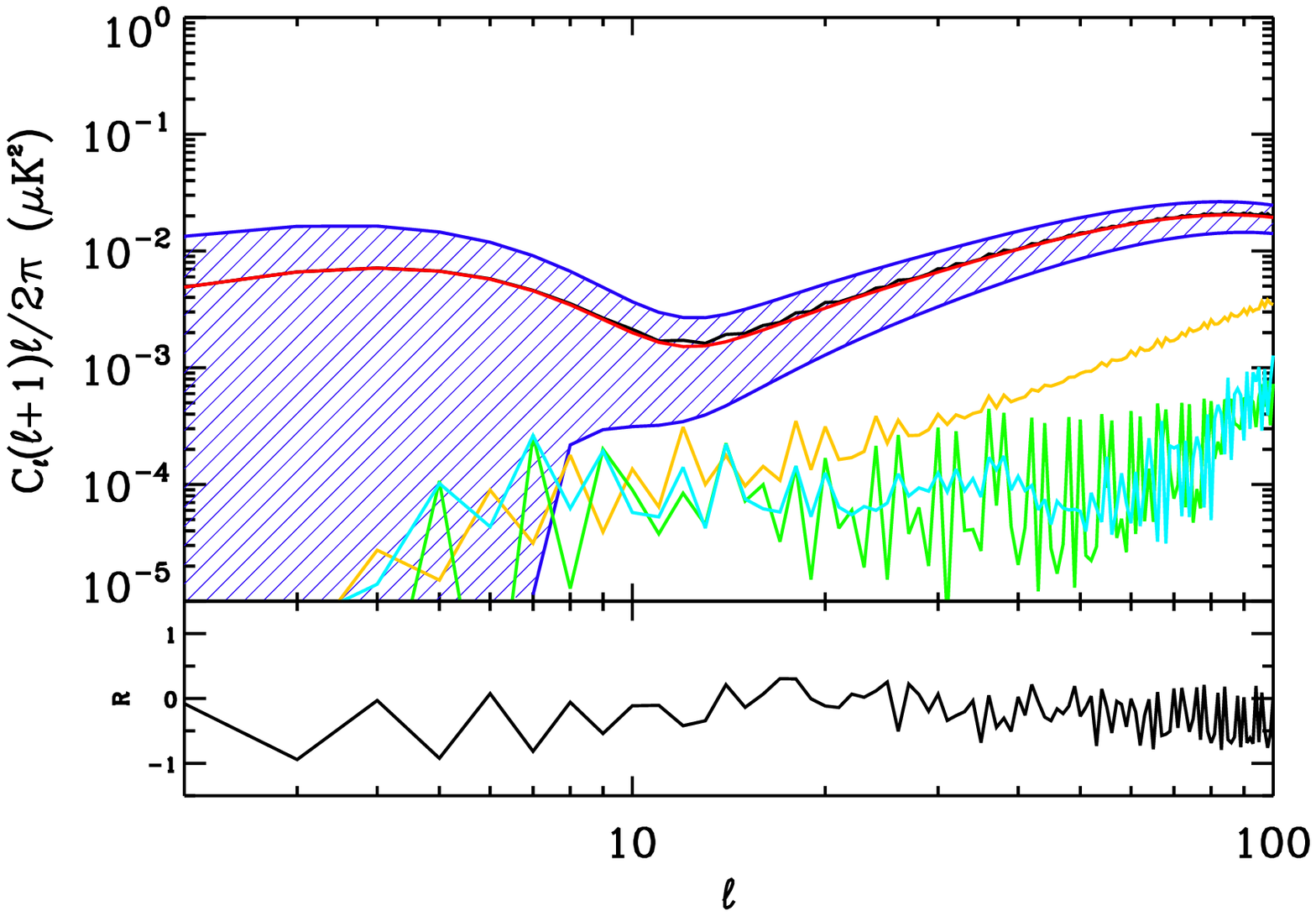,width=5cm}
\psfig{file=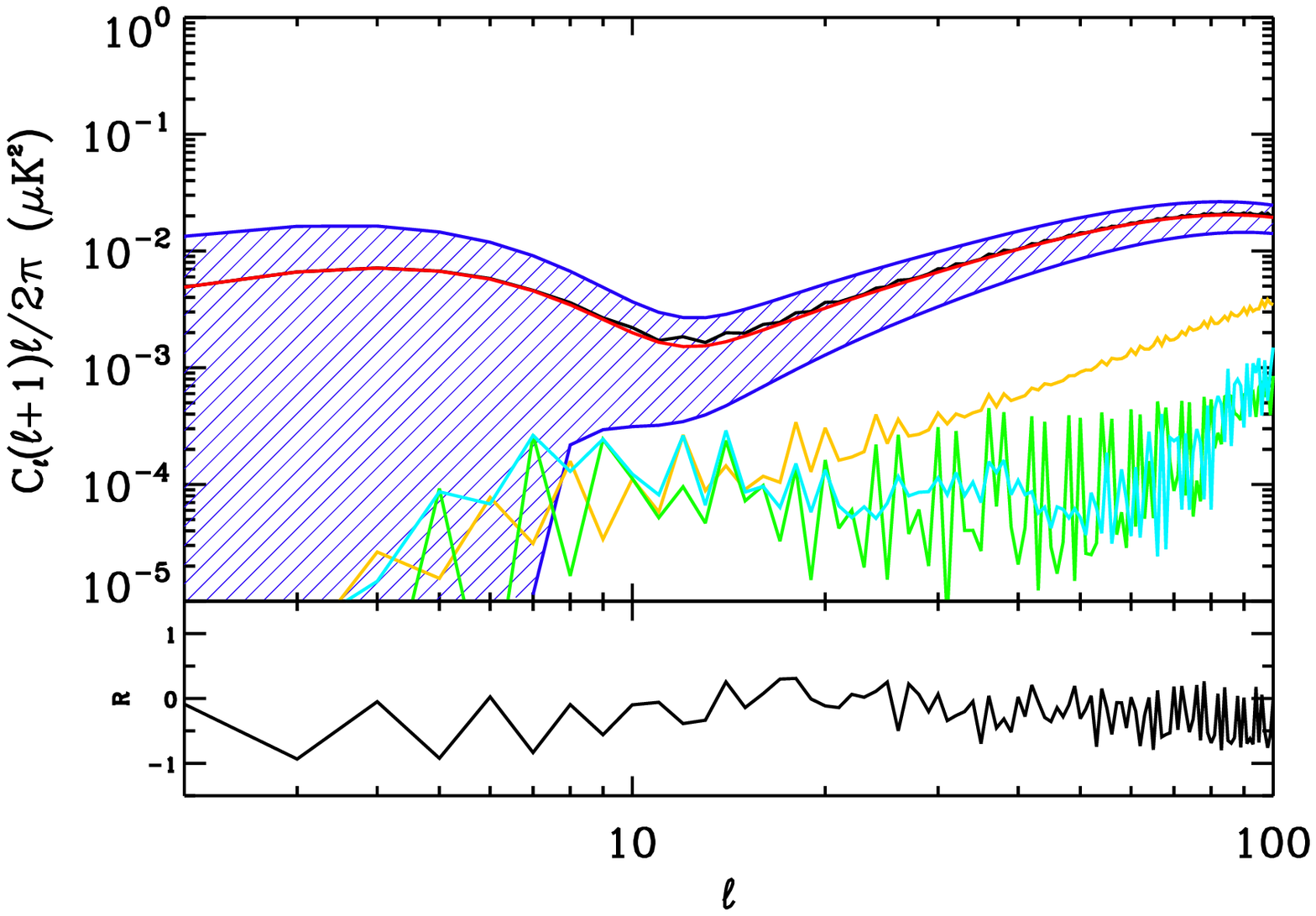,width=5cm}}
\caption{Residual B-mode power spectra for experiments discussed in Table~1.
We show the total residual power spectrum (black), dust residual
(green), synchrotron residual (blue), and detector noise (orange). The
red curve is the primordial B-mode power spectrum with $r=0.3$.  From
top to bottom in each of the rows, the experiments are A to D,
respectively. The first (left) column shows the residual with data
from each of the experiments alone, the middle column is for the
experiments combined with 8-year data from WMAP, and the third (right)
column is for the experiments combined with 14-month data from Planck.
R, at the bottom of each plot, is the cross-correlation of
dust and synchrotron residuals divided by the sum of the
auto-correlations of dust and synchrotron residuals.
}
\label{fig:1}
\end{figure*}

\begin{figure*}[t]
\centerline{\psfig{file=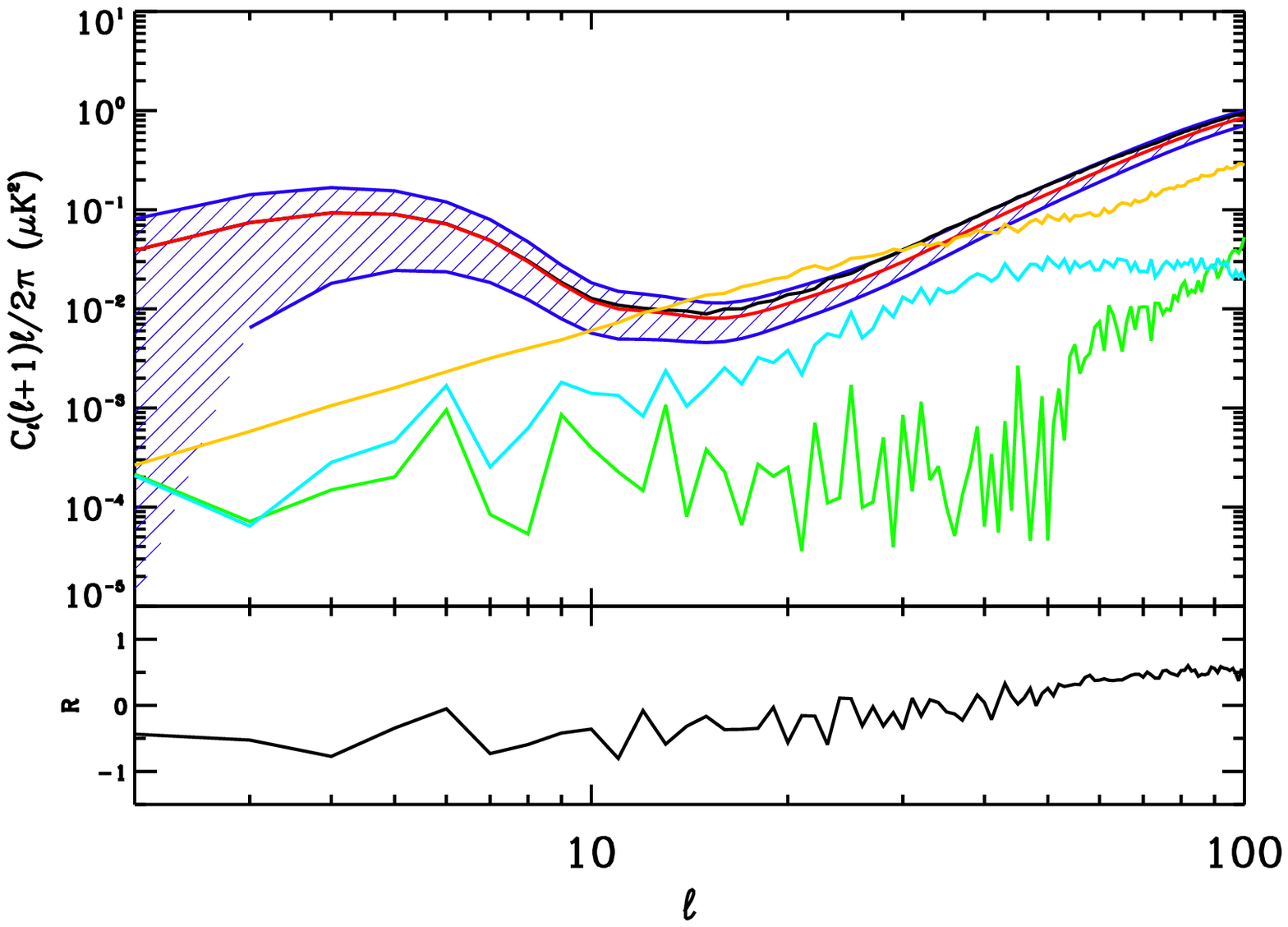,width=5cm}
\psfig{file=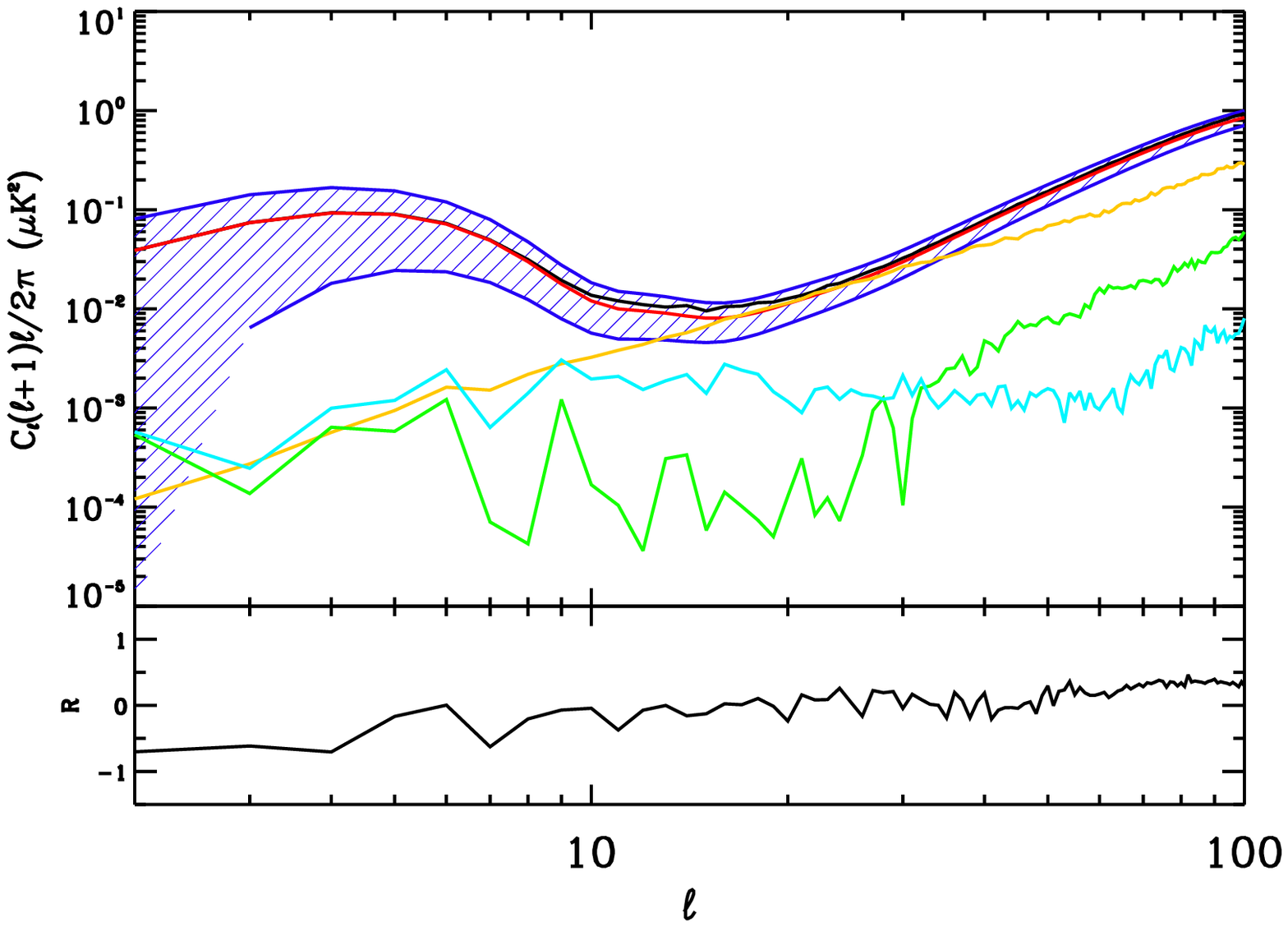,width=5cm}
\psfig{file=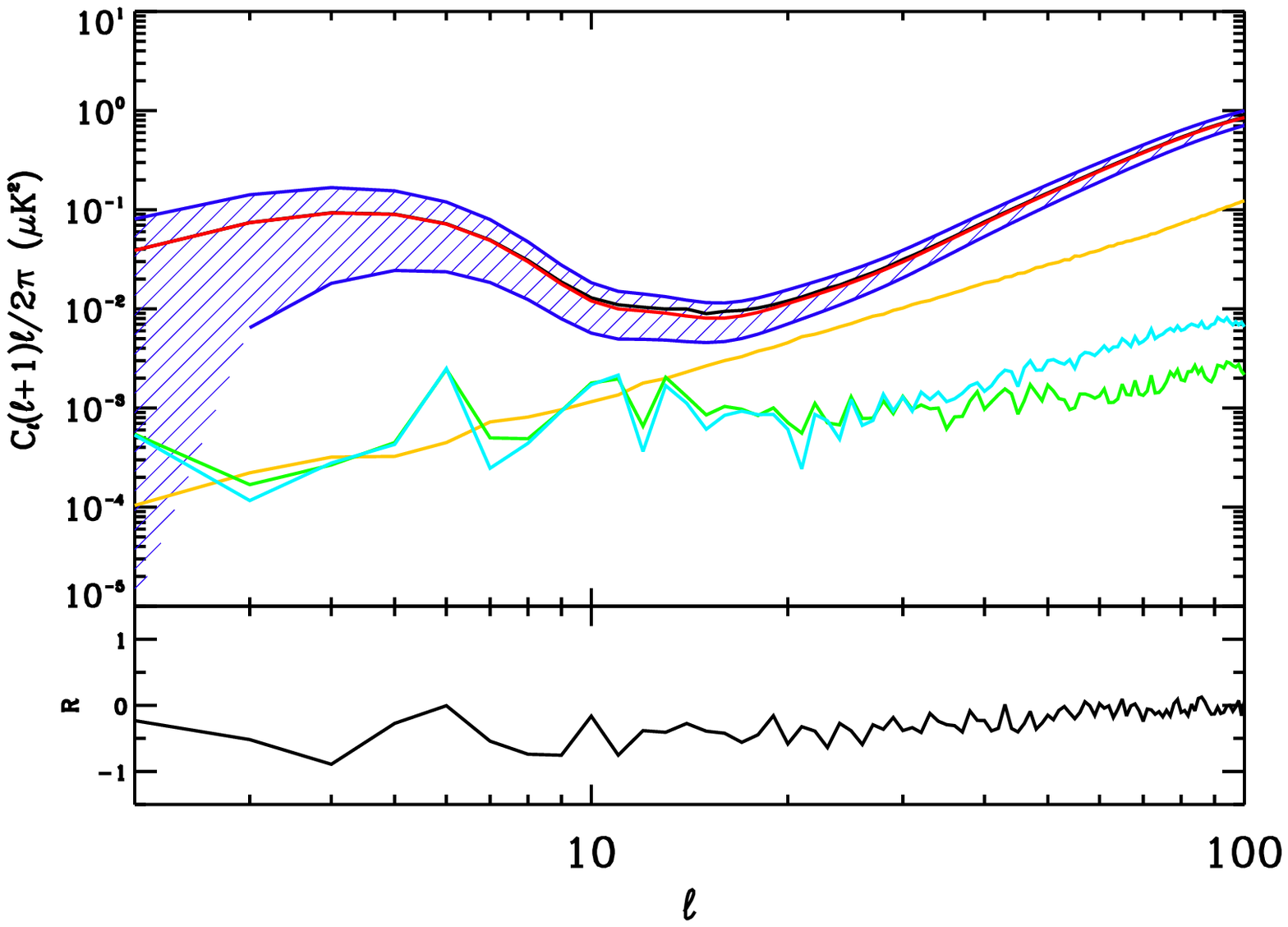,width=5cm}}
\centerline{\psfig{file=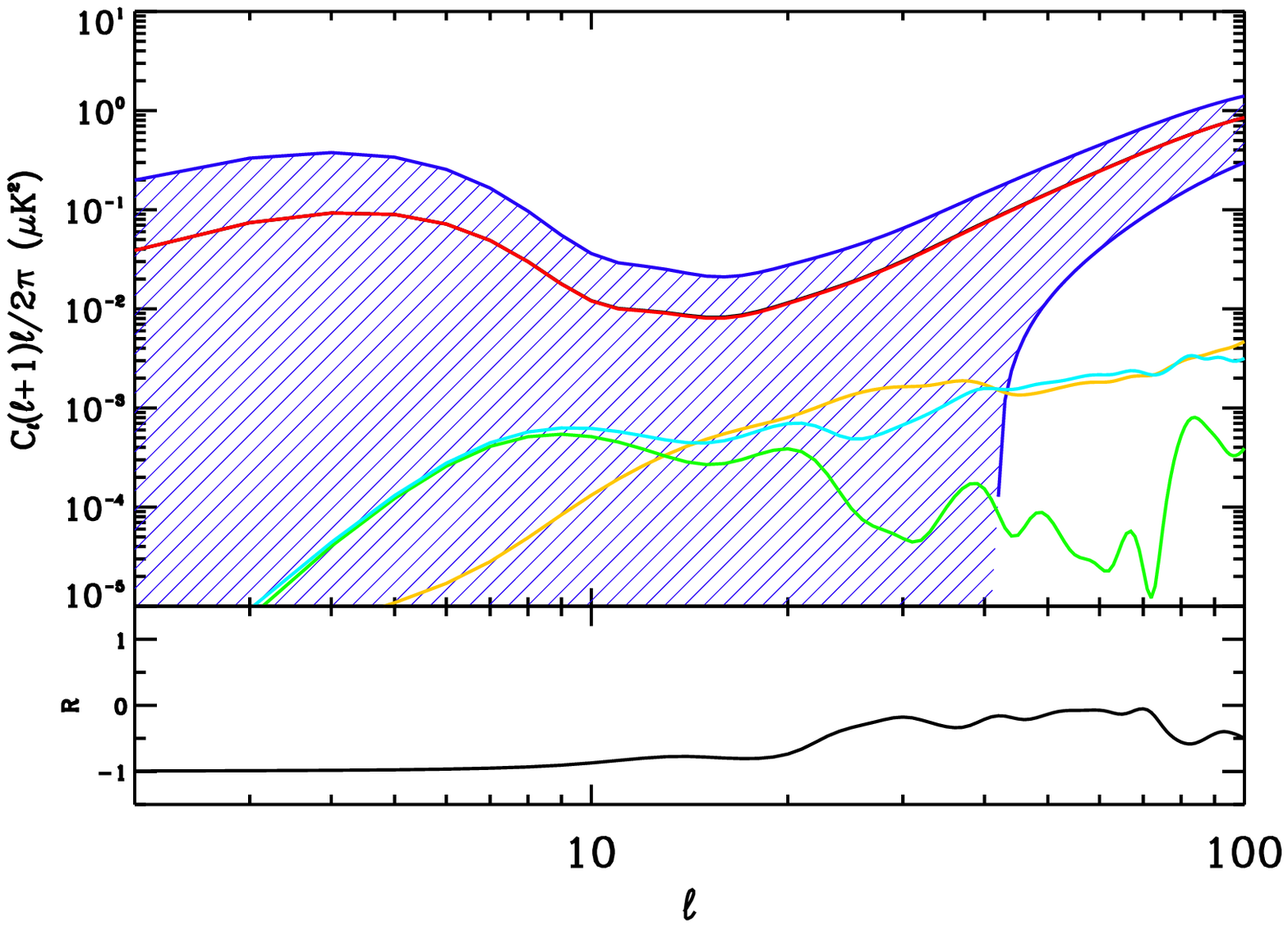,width=5cm}
\psfig{file=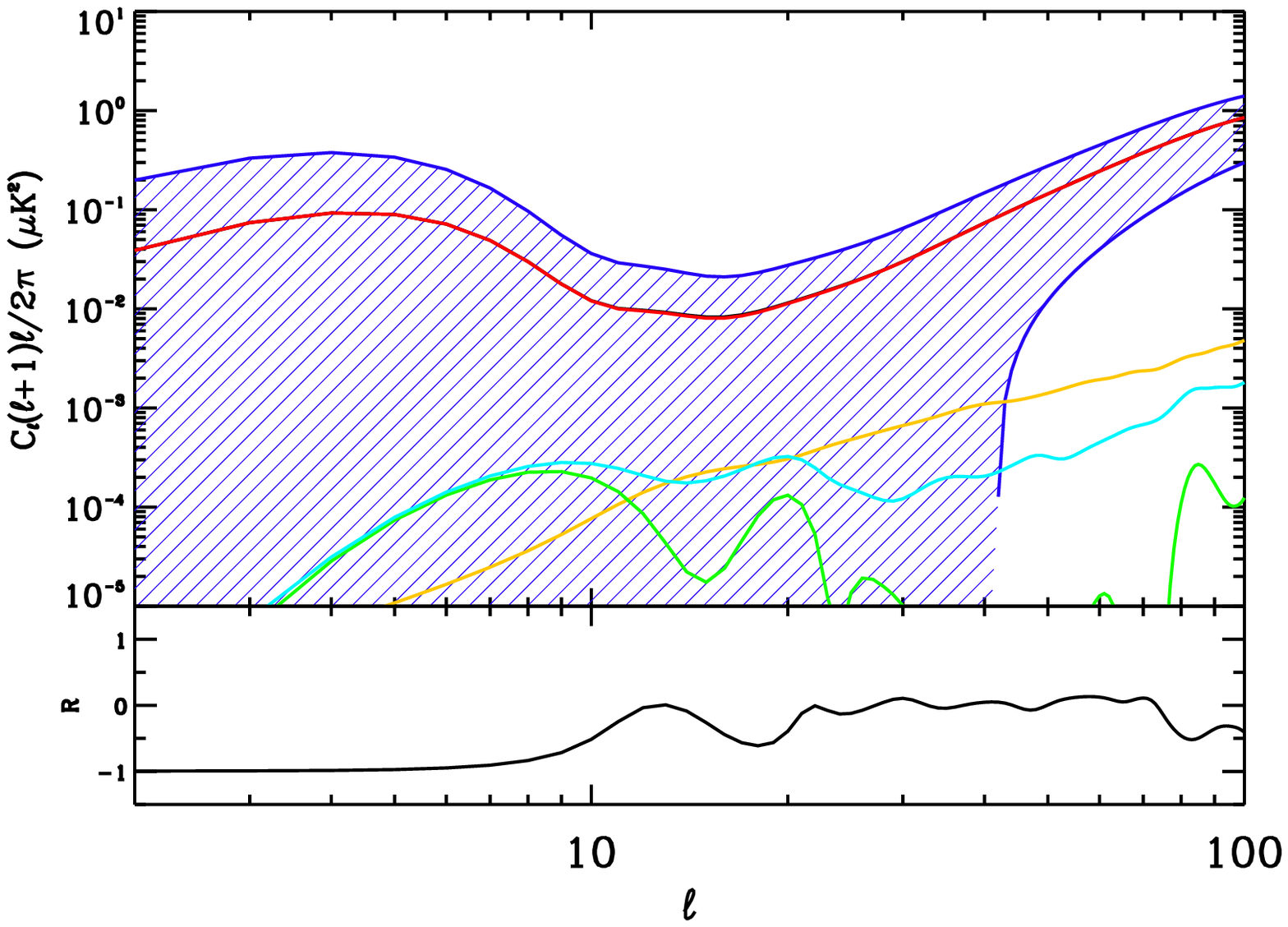,width=5cm}
\psfig{file=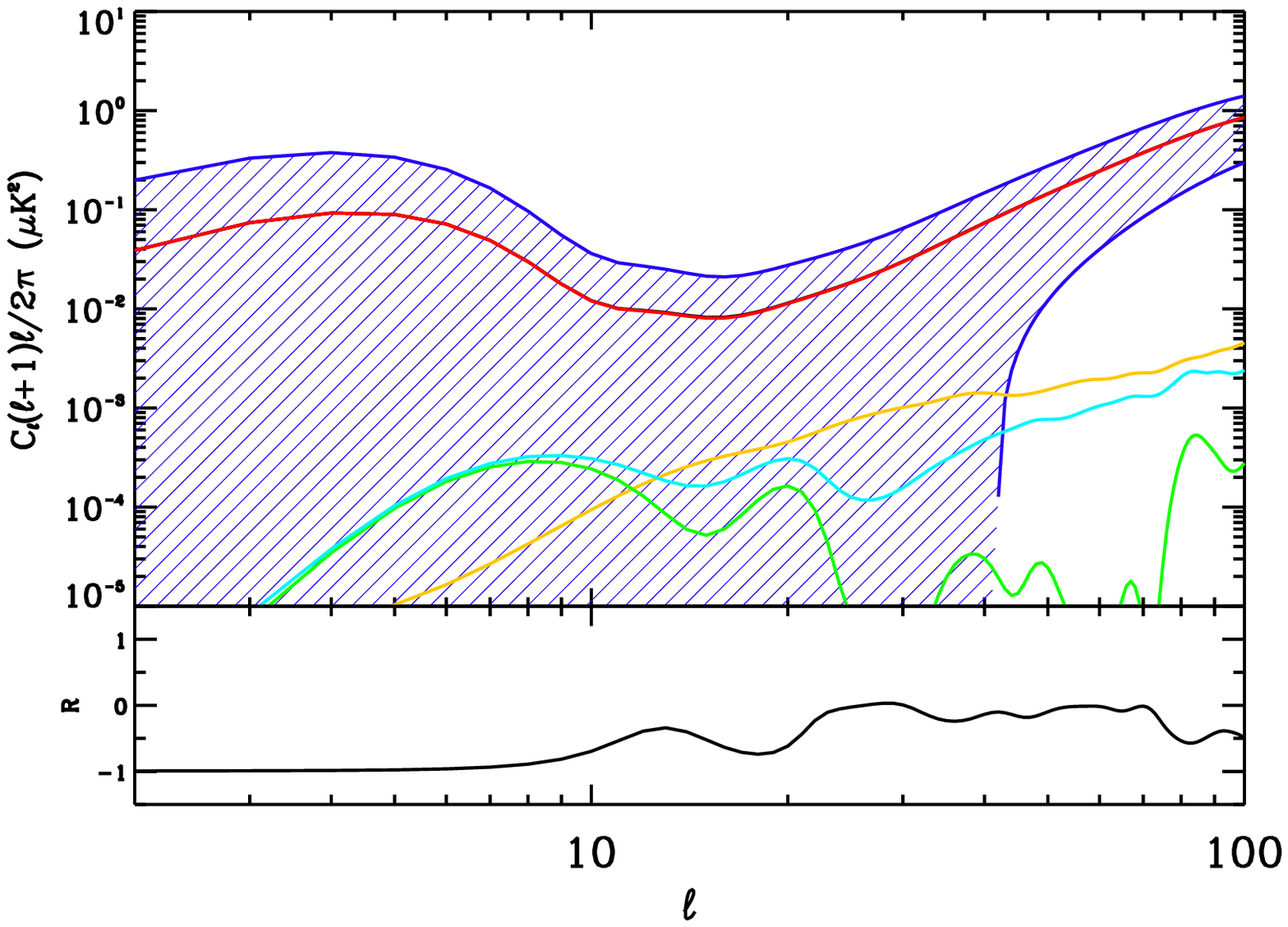,width=5cm}}
\centerline{\psfig{file=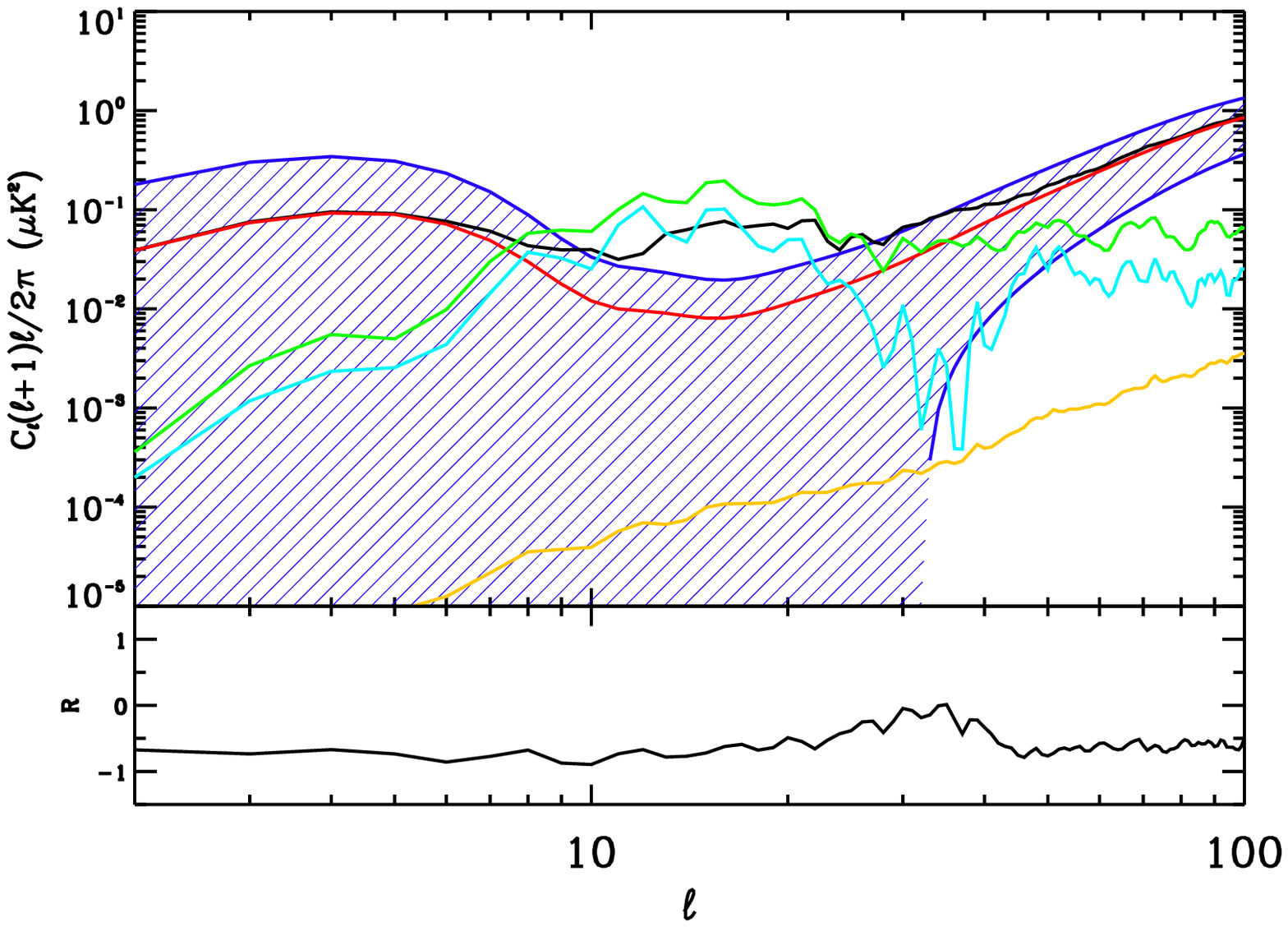,width=5cm}
\psfig{file=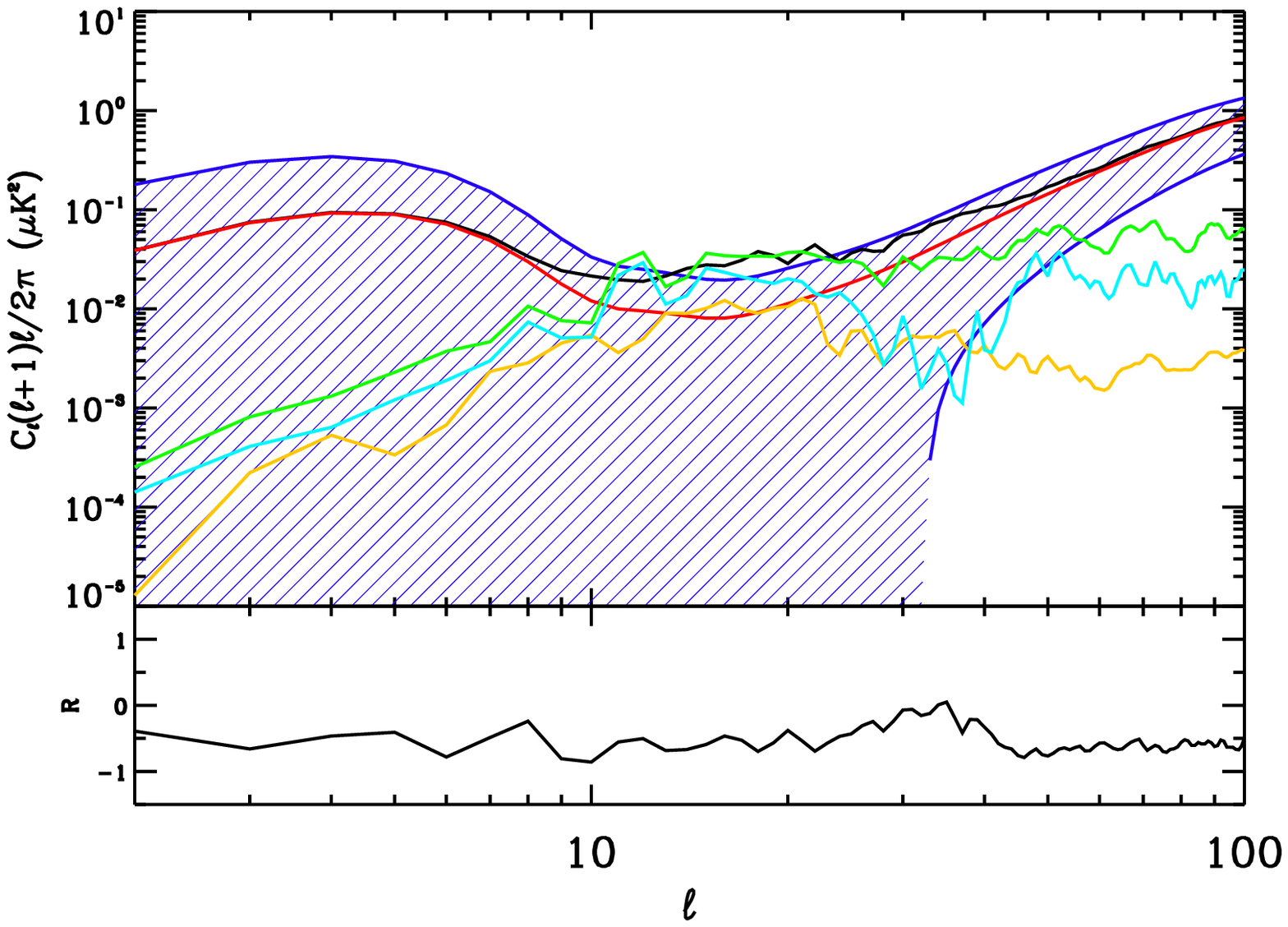,width=5cm}
\psfig{file=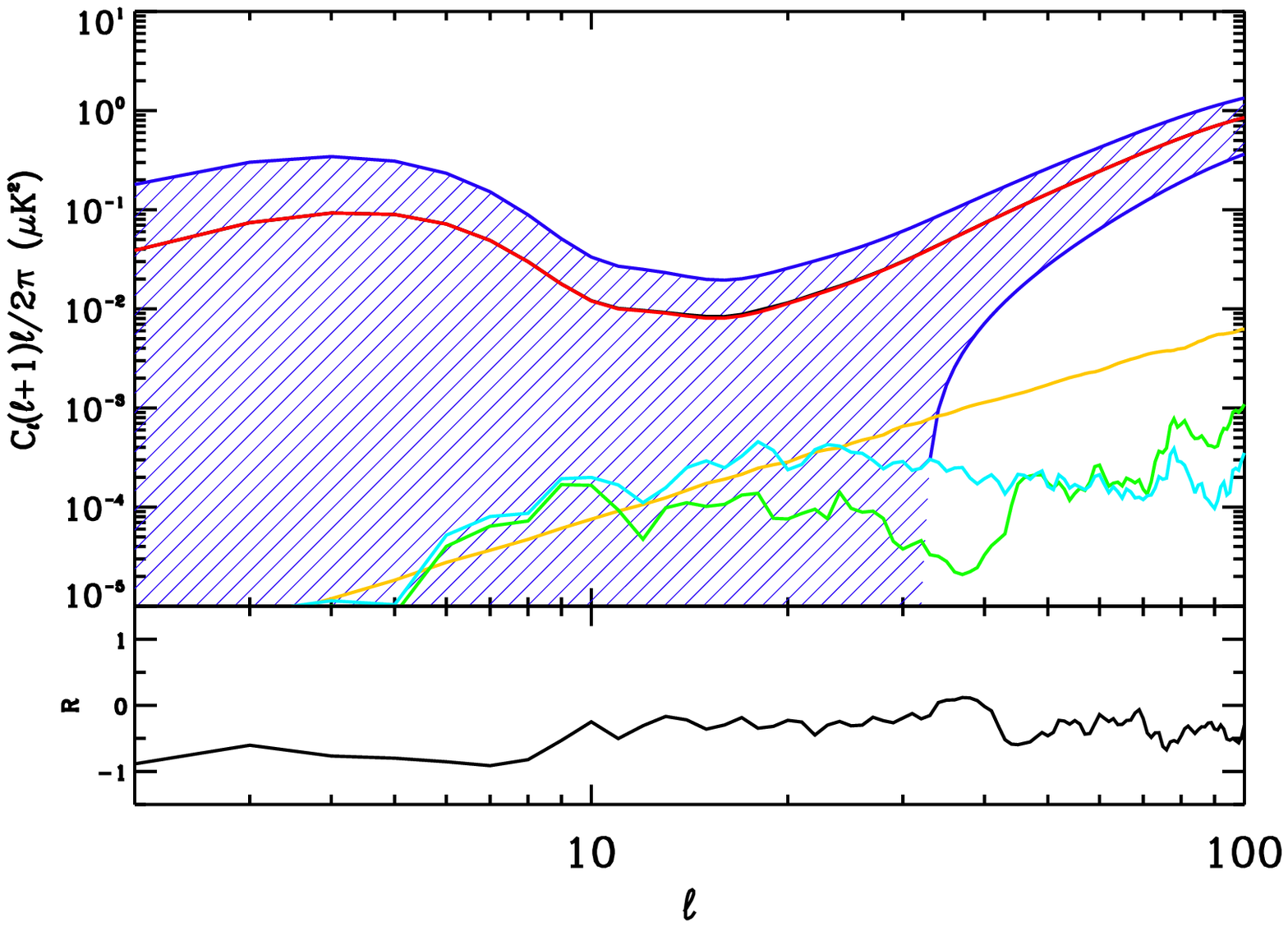,width=5cm}}
\centerline{
\psfig{file=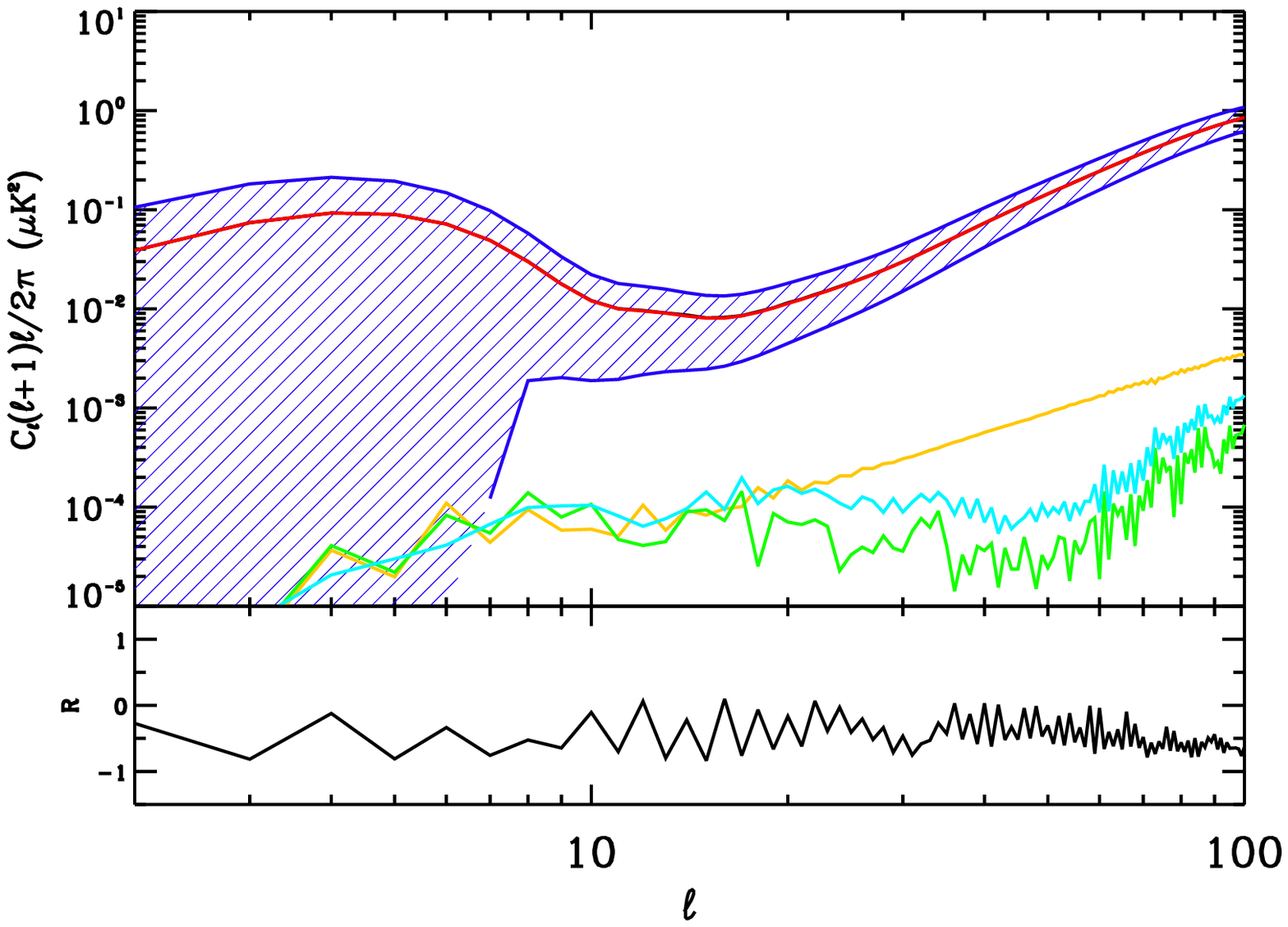,width=5cm}
\psfig{file=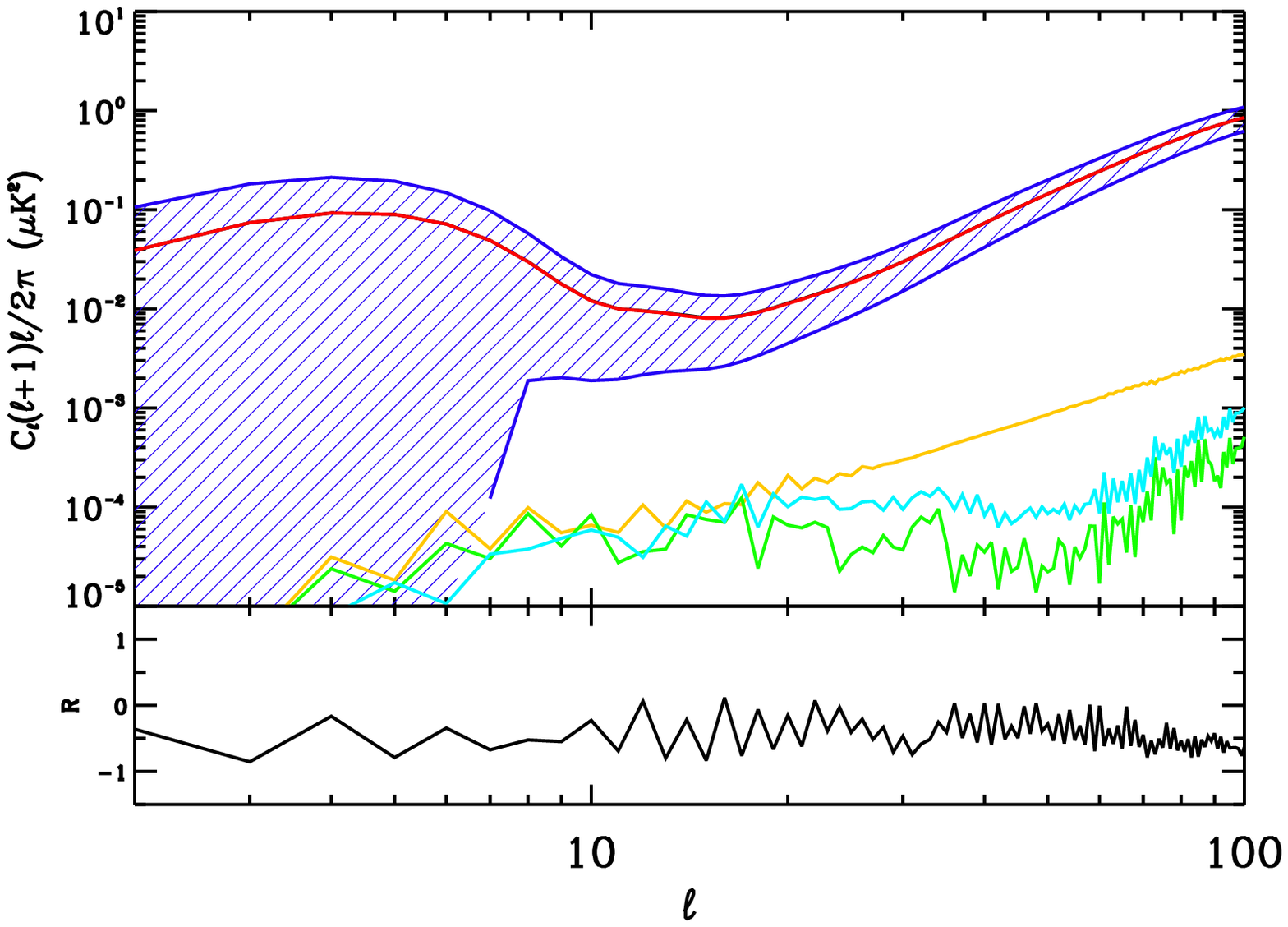,width=5cm}
\psfig{file=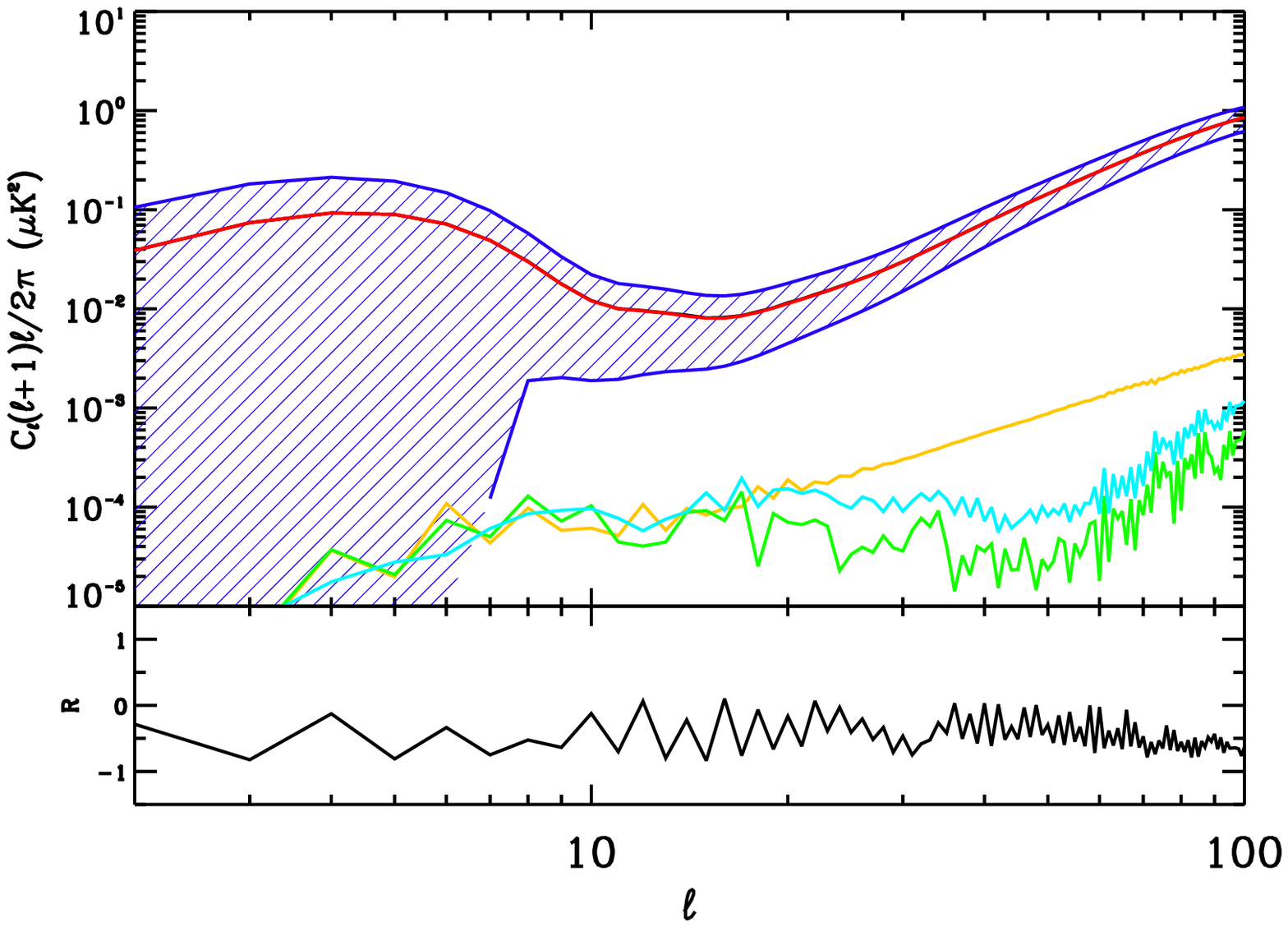,width=5cm}}
\caption{Residual E-mode power spectra for experiments discussed in Table~1.
The curve labels and the panel layout are same as Figure~\ref{fig:1}.}
\label{fig:2}
\end{figure*}

\subsection{CMB Polarization Experiments}
\label{sec:exppa}

We considered hypothetical current and future generation CMB
experiments, and simulated maps for these experiments by adding the
appropriate noise.  Table~1 lists our example experiments. We included
experiments that target the low frequency range (where synchrotron
polarization dominates), the high frequency range (where dust
dominates), and possibilities that span across the frequency range
from 30 GHz to 300 GHz.  Two of the experiments are based on
ground-based attempts to detect primordial B-modes by targeting a
small clean area on the sky (experiments B and C), one suborbital
balloon-borne experiment (experiment A), and another based on a
mission concept for the NASA {\it Inflation Probe}.  Table~1 lists the
frequency bands, the focal-plane sensitivity at each of the frequency
bands, angular resolution assuming a fixed aperture size, the fraction
of sky covered and the duration of the experiment.

When converting simulated foreground and CMB maps to observable maps,
we made certain simplifications.  For example, we did not simulate a
scan pattern on the sky but instead added instrumental noise as white
noise to the maps with uniform sensitivity across the observed area.
In particular in this procedure, we do not account for systematics
such as side-lobes and cross-talks between different modes associated
with non-Gaussian beam shapes \cite{HuZal}. Our aim here is to
estimate the effect of our current knowledge about polarized
foregrounds. Once this issue is better understood we can address
systematics such as those associated with side-lobes and beam shapes.

Our maps are decomposed perfectly to E and B components. In practice,
for partial sky coverage, decomposition of Q and U (Stokes parameters)
maps to E and B polarization modes will lead to mixing between the E
and B modes \cite{Bunn}. This mixing, however, can be reduced through
optimized estimators \cite{Smith} and since these problems are not
exacerbated by foregrounds, we have ignored the added complication in
this study.

\subsection{Foreground removal technique}
\label{sec:clean}

\begin{figure}[h!]
\centerline{
\psfig{file=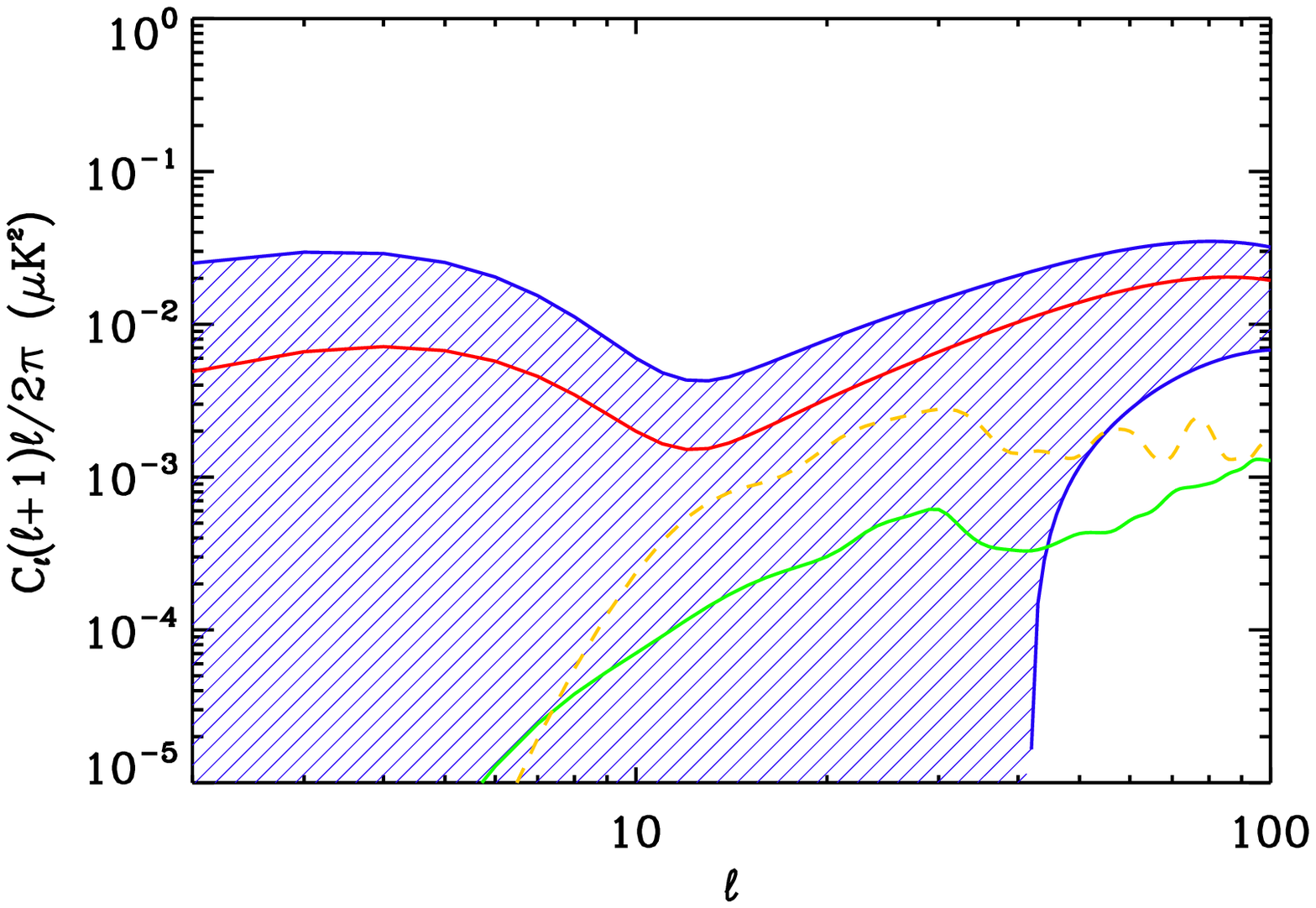,width=8cm}}
\caption{Residual foregrounds obtained using a constant coefficient
for the whole map (dashed orange line) and using the \cite{Tegetal03}
algorithm with one coefficient per $\ell$ mode (solide green line) for
the experiment B setup. The red solid line represents the theoretical
CMB power spectrum for r=0.3 and the blue hatched area the cosmic
variance for the experiment B setup.}
\label{fig:cstcoeftegetal}
\end{figure}

In order to study how well simulated maps for each experiment can be
used to remove foregrounds, we used the cleaning technique outlined in
Ref.~\cite{Tegetal03}, where multifrequency maps from WMAP first-year
data were used to produce the so-called TOH foreground-cleaned CMB
map. This technique allows to take into account the variation of the
spectral index in real space and is more efficient at removing
foregrounds than a simple model with a constant coefficient for the
whole map (all the mode $\ell$) even on small part of the sky like
experiment B coverage (2.36 \% of the sky). When comparing the
residual foreground for a simple coefficient model to the one obtained
by Tegmark {\it et al.}~\cite{Tegetal03} algorithm, we found that the
latter improved the residual foreground level by a factor 5 for
experiment B (see figure \ref{fig:cstcoeftegetal}). 
Therefore we restrain ourselves in the rest of this paper to the
Tegmark {\it et al.}~\cite{Tegetal03} foreground cleaning technique.
The technique recommends taking a linear combination of observed
$a_{\ell m}$'s in each frequency band $i$,
\begin{equation}
a_{\ell m}=\sum_{freq=i}w^i_{\ell}a^i_{\ell m} \, .
\end{equation}
The weights $w_i$ are then to be chosen to minimize foreground
contamination. For polarized observations, we can decompose the signal
at each frequency as
\begin{equation}
a^i_{\ell m} = c_{\ell m} + s^i_{\ell m} + d^i_{\ell m} + n^i_{\ell m} \, ,
\end{equation}
where $c$, $s$, $d$, and $n$ stand respectively for the CMB,
synchrotron, dust, and noise.
We then minimize the resulting power spectrum
\begin{equation}
\langle|a_{\ell m}|^2\rangle= {\bf w_\ell}^T {\cal C} {\bf w_\ell}  \, ,
\label{equ:alm}
\end{equation}
with respect to the weights under the constraint $w_\ell^T \cdot
{\bf e}=1$, when ${\bf e}$ is a column vector of all ones with length
equal to the number of channels. This condition ensures that the CMB
signal is unchanged regardless of the chosen weights. Here, ${\cal 
C}^{ij}_\ell$ matrix represents $\langle (a^i_{\ell m})^\dagger
a^j_{\ell m} \rangle$.  As derived in Ref.~\cite{Tegetal03}, the
weights that minimize the power $\langle|a_{\ell m}|^2\rangle$ are
\begin{equation}
{\bf w_\ell} = \frac{ {\cal C}^{-1} {\bf e}}{e^T {\cal C}^{-1} e} \, .
\end{equation}

\begin{figure*}[t]
\cl{
\psfig{file=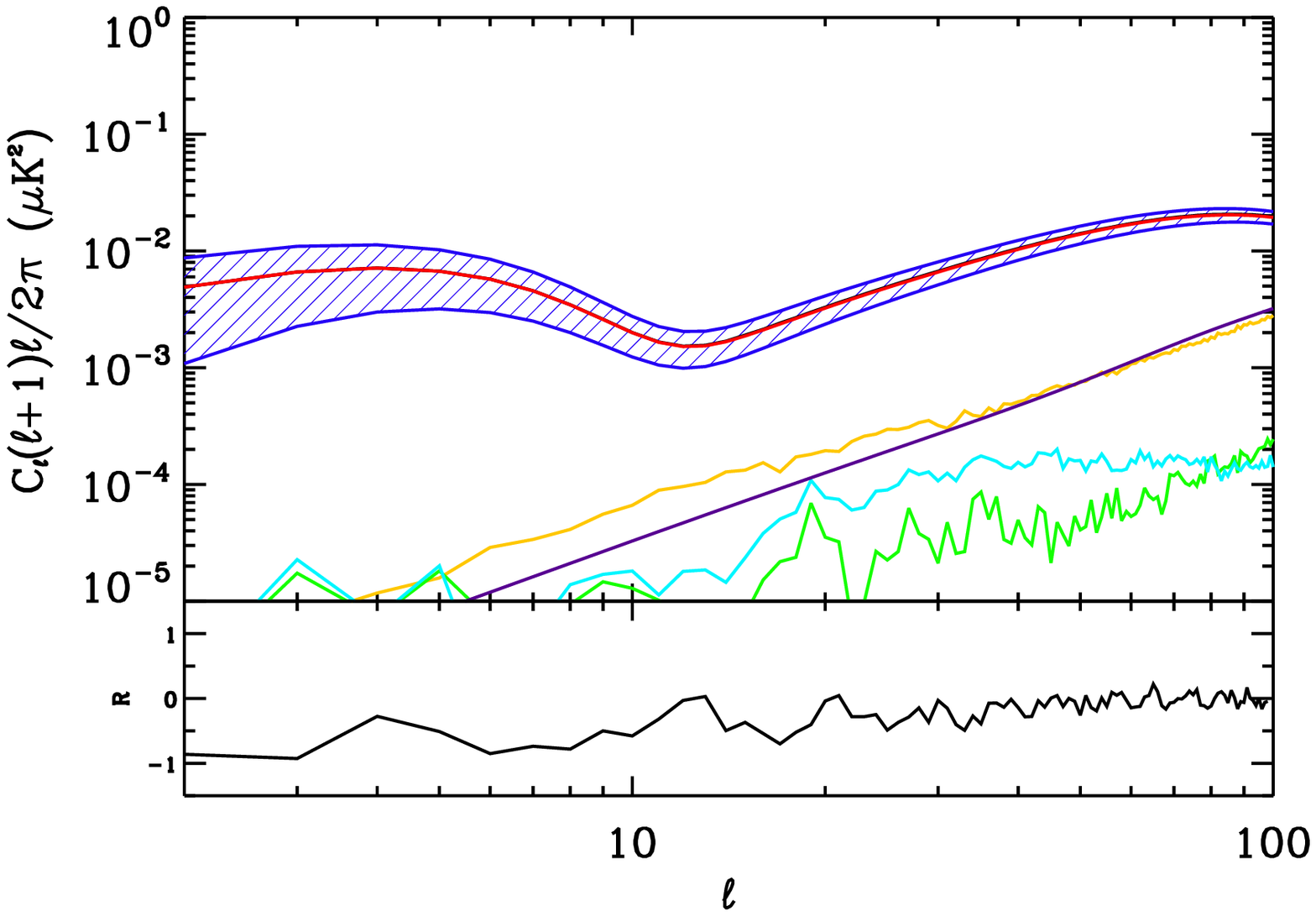,width=7.0cm}\hspace{-1.5cm}
\psfig{file=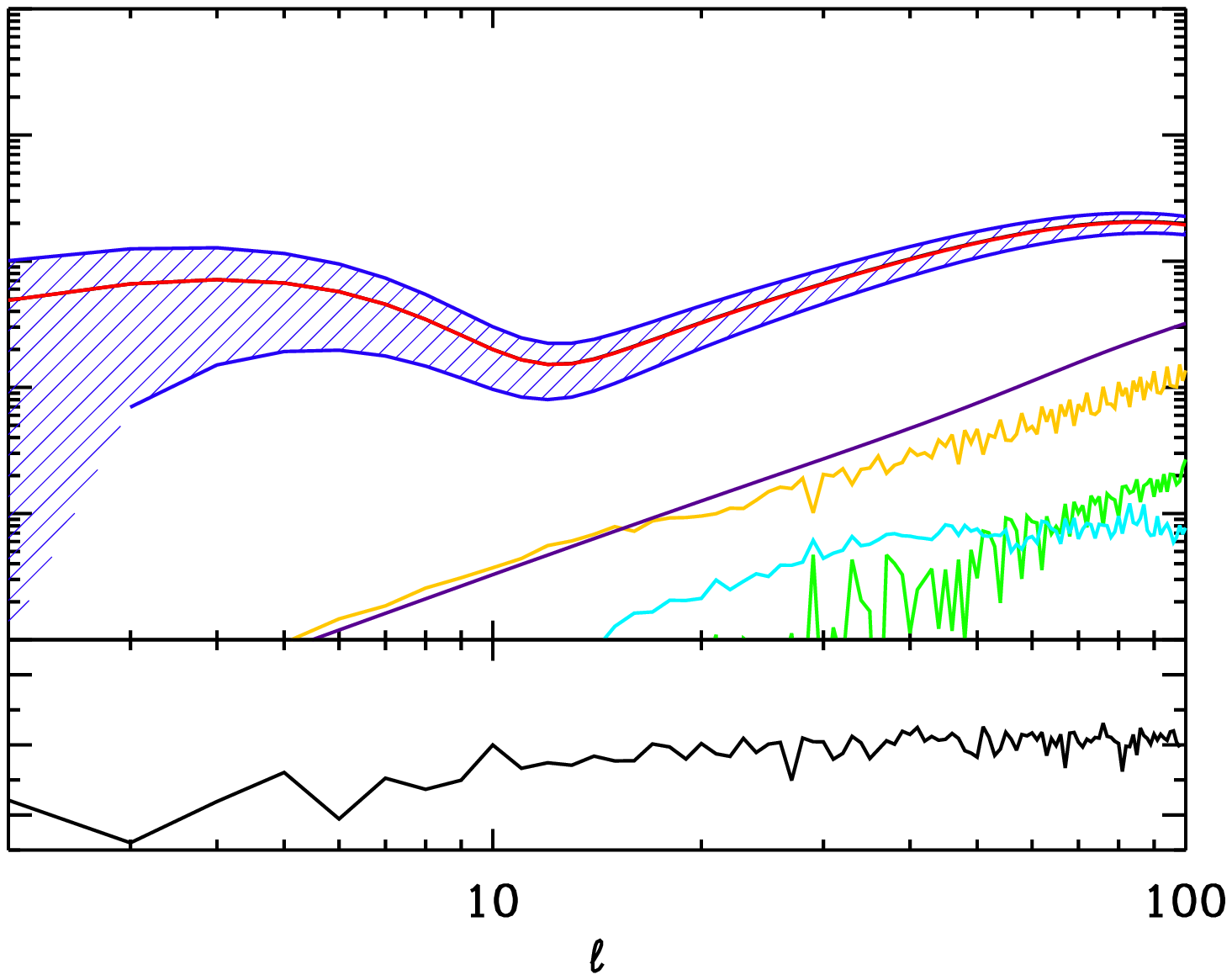,width=7.0cm}\hspace{-1.5cm}
\psfig{file=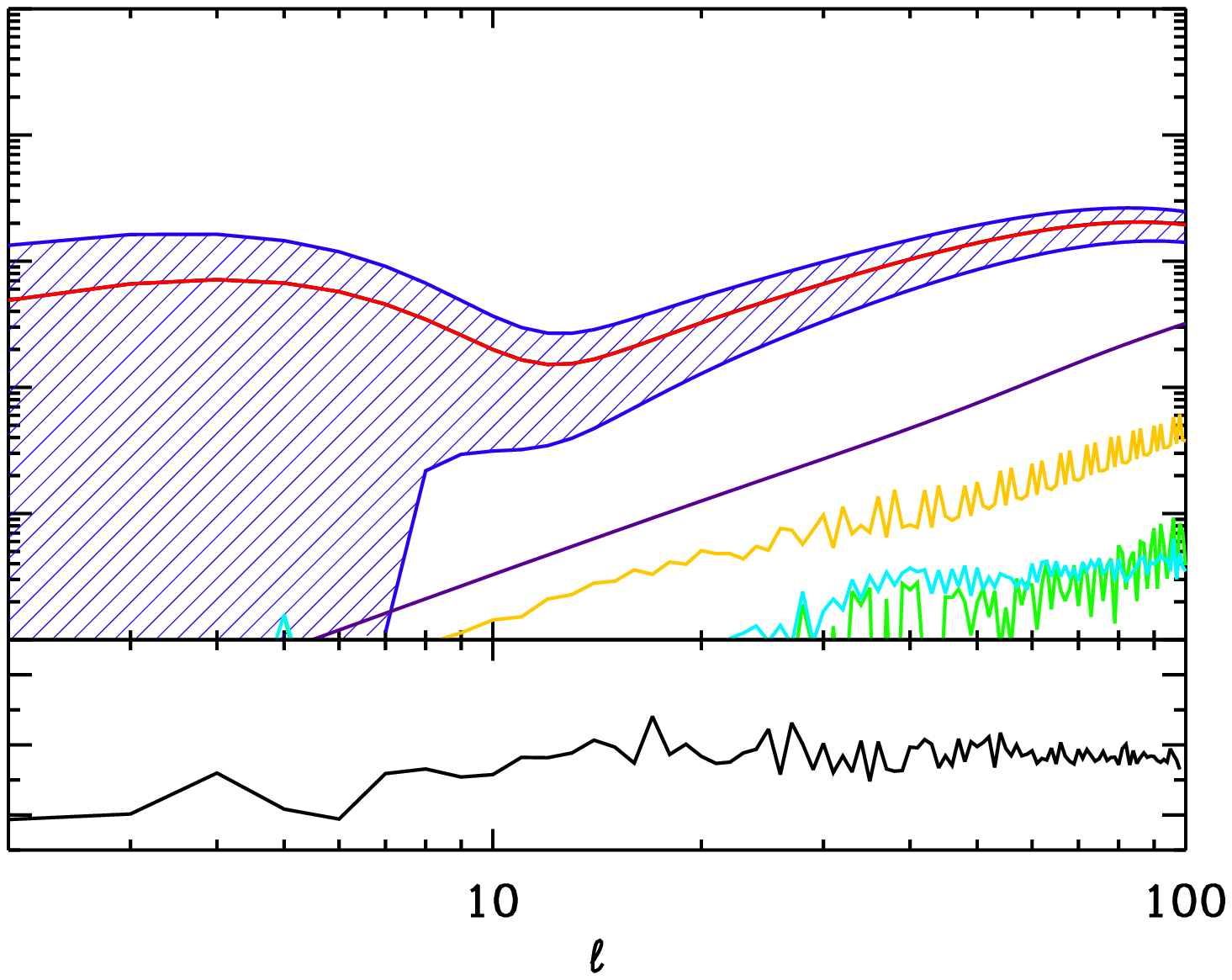,width=7.0cm}
}
\caption{Estimated B mode power spectrum (solid black line) for 
a 1000 detector ``optimized'' experiment covering 65.8, 35.7, 13.4 \%7b
of the sky. The green and light blue solid lines correspond
respectively to the residual dust and synchrotron, the orange solid
line represent the noise power spectrum, the red solid line the
theoretical CMB power spectrum (WMAP cosmological parameters and
r=0.3), and the blue hatched area correspond to the cosmic variance.
R, at the bottom of each plot, is the cross-correlation of
dust and synchrotron residuals divided by the sum of the
auto-correlations of dust and synchrotron residuals.
}
\label{fig:3}
\end{figure*}

\subsection{Minimal Tensor-to-Scalar Ratio}
\label{sec:stat}

For each of the experiments, we compute four quantities to test the
level of residual foreground and detector noise in the cleaned map,
and the achievable lower limit on the tensor-to-scalar ratio $r$,
which we quote as a 3 $\sigma$ confidence limit. 

1. {\bf Average noise power}
in the map (where $N_\ell$ is the noise power spectrum) between $\ell$
of 2 and 100 :
\begin{equation}
\mathrm{av\_noise} =
\frac{1}{99}\sum_{\ell=2}^{100}{{N_\ell(\ell+1)\ell \over 2\pi}} \,.
\end{equation}

2. {\bf Average noise power spectrum variance} 
of the map between $\ell$ of 2 and 100 :
\begin{equation}
\mathrm{av\_noise\_var} =
\frac{1}{99}\sum_{\ell=2}^{100}{{N_\ell(\ell+1)\ell \over
    2\pi}\sqrt{2\over (2\ell+1)f_{sky}}} \,.
\end{equation}

3. {\bf Average residual in the estimated power spectrum} 
of the map after subtracting the CMB and noise power spectrum
between $\ell$ of 2 and 100: 
\begin{equation}
\mathrm{av\_fgd\_var} =
\frac{1}{99}\sum_{\ell=2}^{100}{{R_\ell(\ell+1)\ell \over 2\pi}} \,.
\end{equation}

4. {\bf $r_g$, ``Gaussian'' estimate of achievable tensor to scalar
ratio}, the foreground residual $R_\ell$ is assumed to be an extra
Gaussian noise and we sum over all the $\ell$ modes optimally to
constrain the overall signal-to-noise ratio to be above 3 for $r_g$ :
\begin{equation}
\sqrt{\sum_{\ell=\ell_{min}}^{\ell=\ell_{max}}\left(C_\ell(r_g) \over
(R_\ell + (N_\ell + C_\ell(r_g))\sqrt{2\over (2\ell +1)
f_{sky}})\right)^2}=3
\end{equation}
When not specified, $\ell_{\rm min}$ is given by the largest mode that
fits within the sky area observed by a given experiment, or
$\pi/\theta$, where $\theta$ is the small side of the survey (for
rectangular regions) in radians. We stress that this estimate may be
misleading because treating the residual as gaussian distributed noise
is incorrect. We provide this estimate here as a comparison to the
estimate that is described next.

5. {\bf $r$, estimate of achievable tensor to scalar
  ratio } computed such that given the signal
$s=\sum_{\ell=\ell_{\rm min}}^{\ell_{\rm
    max}}{C_\ell(\ell+1)\ell p_\ell / 2\pi}$ and the residual
$u=\sum_{\ell=\ell_{\rm min}}^{\ell_{\rm max}}{R_\ell(\ell+1)\ell p_\ell / 2\pi}$, 
\begin{equation}
\int_{u}^{\infty}{P(s|r_{min}) {\rm d}s}= 0.99
\label{equ:dis99}
\end{equation}
$p_\ell$ are the weights used to sum up the different $\ell$ modes.
This $r$ corresponds to a tensor-to-scalar ratio such that given a
certain experimental noise and foreground residual, the signal is 99\%
likely to be above the foreground residual. We note that this
definition is somewhat arbitrary in that we have defined detection as
signal over expected residual is greater than unity. A correct
estimate would require that we quantify the likelihood of the
parameters used to create the foreground maps.  This is beyond the
scope of our introductory study.

In equation~(\ref{equ:dis99}) 
$P(s|r)$ is multi-variate Gaussian in $a_{\ell m}$. Instead of
generating Monte-Carlo simulations and using the resulting $P(s|r)$ in
equation~\ref{equ:dis99}, we employ an approximation. To do so, we
first estimate the variance of $s$. This is easily computed by
considering real independently distributed Gaussian variables $x_k$  
in terms of which
we have $s=\sum_kx_k^2 - \sum_\ell \ell(\ell+1)N_\ell/2\pi$. The
variance of $s$, denoted by $\sigma_s^2$ is then trivially computed
using $\langle x_k^4 \rangle = 3\langle x_k^2\rangle^2$  
to give $\sigma_s^2 = \sum_k 2 \langle x_k^2 \rangle^2$. Written in
terms of the zero-residual map power spectrum, we have 
\begin{equation}
\sigma_s^2=\sum_{\ell=\ell_{\rm lmin}}^{\ell_{\rm
max}}{\left((C_\ell+N_\ell)(\ell+1)\ell p_\ell\over 2\pi\right)^2{2\over(2\ell+1)f_{sky}}}\,.
\label{equ:sigs}
\end{equation}

One can verify that the central limit theorem applies for the
distributions under consideration by ascertaining that the Lyapunov
condition holds.  Given the large number of modes we sum over (even
for $\ell_{\rm max} = 20$) we expect $P(s|r)$ to be approximately
Gaussian (close to the peak) with variance given by $\sigma_s^2$. In
order to see if this is so in all the region required for the
integration in equation~\ref{equ:dis99}, we generated Monte Carlo
realizations of $s$ to determine $P(s|r)$. We found that a Gaussian
with variance $\sigma_s^2$ fits $P(s|r)$ very well.  To compute the
limiting tensor to scalar ratio we then simply set $s-2.32\sigma_s$
equal to the sum of the residuals $u$.  We obtain our minimum
achievable $r$ by optimizing the $\ell$ mode sum using the weight
$p_\ell$ (these weights are determined numerically by minimizing
$r$). We note that this minimization procedure is not crucial. Simple
high and low-$\ell$ band power estimates give results that are
consistent with the above scheme. The separation into high and
low-$\ell$ band-powers is motivated by the distinct contributions to
the primordial B-mode power spectrum from the recombination and
reionization epochs. A definitive detection of the primordial B mode
signal in the future will have to rely on the knowledge of the
primordial B-mode power spectrum and consistent estimates of $r$ from
the low and high-$\ell$ regions. This requires that we understand the
reionization history well using the large angle E mode signal
\cite{Kaplinghat02}.

\begin{figure*}[t]
\cl{
\psfig{file=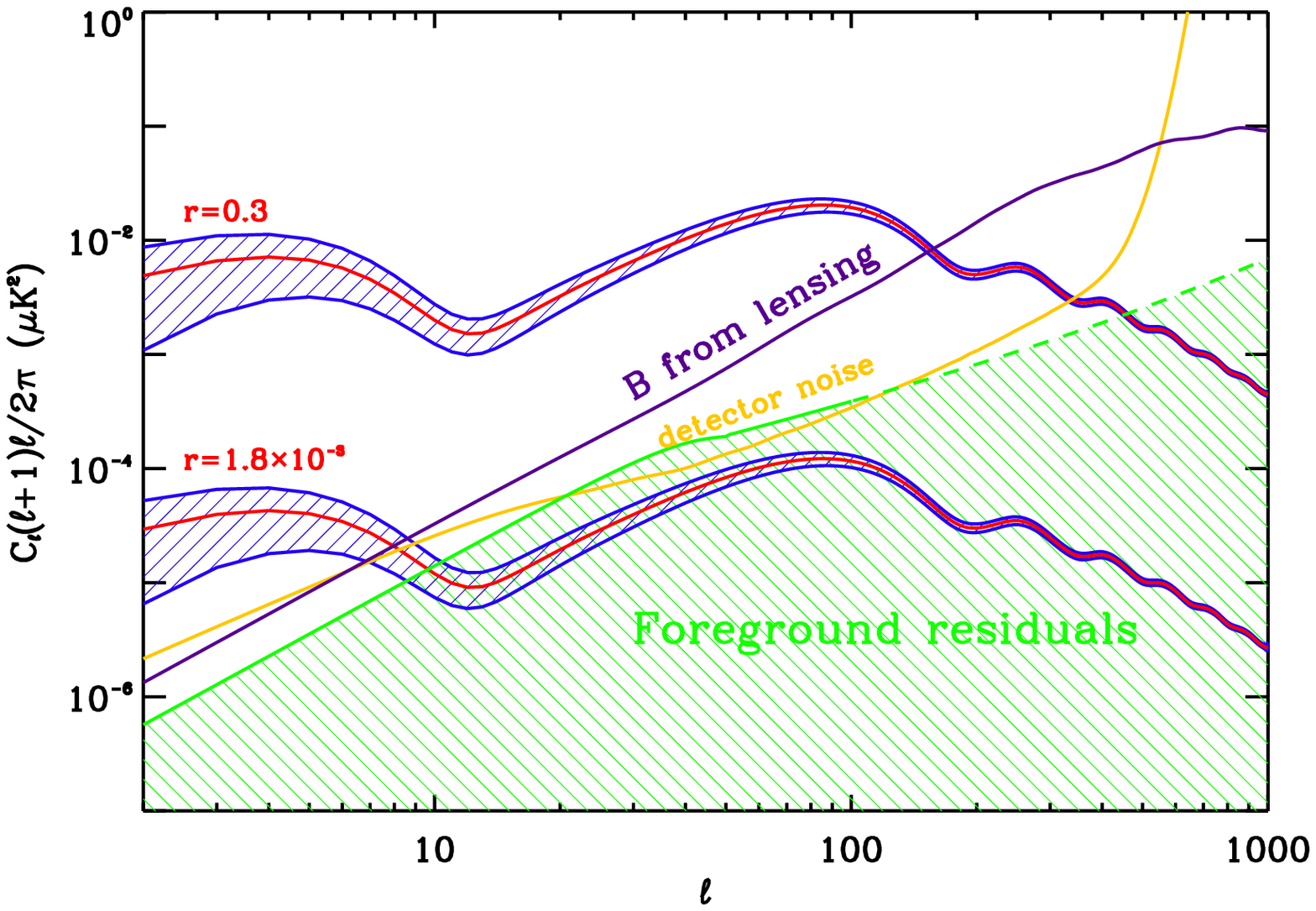,width=17cm}
}
\caption{The detectable range of the tensor-to-scalar ratio with a
next generation CMB experiment whose frequency band selection is
optimized following the calculation presented here.  The lower and
upper curves for primordial B-modes show an amplitude with a
tensor-to-scalar ratio of 0.3 and $1.8 \times 10^{-3}$ respectively,
roughly indicating the current upper limit on $r$ and the limit
reachable with this experiment. The green hatched area is the region
where residual foreground noise dominates, due to both polarized dust
and synchrotron.  The detector noise shows the variance with
$\sqrt{2/(2\ell+1)/f_{sky}}N_\ell^{\rm noise}$.  The curve labeled ``B from
lensing'' is the lensing generated B-mode power spectrum which acts as
an additional source of irreducible noise. We assume a survey of 66\%
of the sky over 2 years.} 
\end{figure*}

\subsection{Results}

We performed the foreground cleaning technique outlined in
Section~\ref{sec:clean} on the four experiments tabulated in
Table~\ref{tab:exppa} and described in Section~\ref{sec:exppa}.  We
considered both E and B mode power spectrum measurements. Since the
main template polarization map from WMAP at 23 GHz has a resolution of
about 1 degree, we limited our discussion to angular scale between
$\ell=2$ and 100.  With higher resolution maps, such as in a few years
from {\it Planck}, our procedure can be easily extended to account for
a wider multipole range. Limiting our discussion to a multipole of
about 100 is adequate since the primordial tensor mode power spectrum
in CMB B-modes has two significant bumps at 
$\ell$ of 10, associated with reionization scattering, and at $\ell$
of 100, associated with the horizon size at the matter-radiation
equality \cite{Pritchard}.  The ongoing and planned experiments target
these two bumps, though ground-based observations are mostly
restricted to the bump at $\ell$ of 100 due to large cosmic variance
at low multipoles and large scale systematics due to atmospheric
emission. 

In terms of the experiments outlined in Table~\ref{tab:exppa},
experiment C performed worse than the other options as only two
frequencies are covered and the foregrounds, on average, dominate over
primordial CMB power at the frequencies selected for observations.
For experiment C, the main contaminant is the residual dust at 90 GHz
(see, figures \ref{fig:1} and \ref{fig:2}).  Experiment A, while
covering a wide range of frequencies, suffers from a low
signal-to-noise in each of the channels as the experiment targets a
wider area in 10 days. This low signal-to-noise ratio only allows
foregrounds to be removed adequately at low multipoles, while at a
multipole of 100, detector noise at each of the frequencies starts to
dominate and no foreground discrimination is possible. As can be seen
from Table~\ref{tab:exppa}, experiment D with a wide range of
frequency coverage and extremely high sensitivity, easily separates
foregrounds over the multipole range of interest. This kind of
frequency coverage and high sensitivity is achievable from space.

The discussion, so far, has focused on the ability of each experiment
to remove and reduce the foreground contamination. However, there is
already a good template for the synchrotron emission on large angular
scales at 23 GHz from WMAP. Thus, one can improve the removal further
by making use of polarization maps from WMAP. To study this we include
WMAP as additional channels. While other multifrequency maps are
available, what is important for this analysis is the low-frequency
anchor provided by WMAP at 23 GHz. To further improve the sensitivity,
we added WMAP data assuming 8 years of observation.  

As shown in the middle column of Figs.~\ref{fig:1} and \ref{fig:2},
while the recovered power spectrum improve for experiments C, the
residual foreground level still dominates the $r=0.3$ CMB B-mode power
on angular scales of a few degrees owing to the limited B mode
polarization reach of WMAP even with 8 years of data.  Note that when
adding WMAP8 (or Planck below) to experiments A-D, we are only using
WMAP8 or Planck to help clean foregrounds and add to the signal 
{\it in the patch of the sky observed by experiments A-D}.

In mid-2008, {\it Planck} will be launched and will make CMB
polarization measurements over a 14 month period over the whole
sky. The polarization maps are expected to have sensitivities roughly
a factor of 10 better than 8-year WMAP data \cite{Planck06}.
Furthermore, Planck will provide anchors at both low- and
high-frequency ends tracing both synchrotron and dust, respectively.
With 14-month polarization data, and using both the lowest and highest
frequency bands as tracers of foreground polarization, we computed the
residual foreground level for the same four experiments. 
We include all the Planck channels (HFI and LFI) in our
  analysis.
The results are summarized in the right column of Figs.~\ref{fig:1}
and \ref{fig:2}.  As shown in these figures, Planck data improves
foreground removal significantly compared to WMAP. This is because
Planck has information at higher frequencies than WMAP, say at 150 GHz
and above, where dust polarization dominates with Planck capturing
that information out to a frequency of 353 GHz.

An experiment such as C which is limited by residual dust at 90 GHz,
when combined with Planck, can limit the tensor-to-scalar ratio to be
below 0.02 at the 99\% confidence level, as shown by the $r$ estimate.
Similarly, a ground-based experiment such as B which is limited by
residual synchrotron at low frequencies can be combined with WMAP
8-year data to limit the tensor-to-scalar ratio to be 0.026 at the
99\% confidence level.  Note that this limit is better than that for
the combination of experiment B and Planck which leads to 0.035 at the
99\% confidence level.  This traces primarily to the fact that while
the low noise channels of WMAP 8-year data are effective at removing
synchrotron foreground, the high frequency channels of experiment B
(though low noise) is not at adequate for an improved removal of
synchrotron when combined with Planck.  Just as the combination of B
and WMAP8 is better, the combination of C and Planck is also better
due to differences in the frequency coverage.  In general, when we
compare experiments A, B or C alone, plus Planck and with Planck
alone, there is an improvement of a factor between 2 and 40 in the
tensor-to-scalar ratio limit. This shows that ground-based experiments
and Planck complement each other, the former by providing a CMB
channel with very low noise level, the latter by providing good
foreground estimates.

Note that, however, these limits are a factor of 2 to 4 within the
published target goal 0.01 of the {\it Inflation Probe}
\cite{Weiss}. Our study thus shows that the limit of 0.02 to 0.04 at
the 99\% confidence may be achieved from the ground with experiments
that target an area about or slightly less than 3\% of the sky with
two to three-year integration to improve sensitivity, and with the
help of either WMAP 8-year or Planck data depending on the frequency
range of the ground-based experiment.  Our study further shows that it
is unlikely that any ground-based experiment, when combined with
either WMAP8 or Planck, will reach the published goal of 0.01 at the
99\% confidence level. The two experiments B and C we have considered
are long-term goals of some of the existing experiments. Thus, these
capture what is to be expected after Planck is launched and involve
significant number of detectors in the focal plane integrating for a
long time relative to some of the on going CMB experiments. Thus, it
is reasonable for us to state that to reach the published goal of 0.01
within a year of observations, one must consider observations from
space.

\begin{table*}
{\scriptsize
\begin{center}
\begin{tabular}{cccccc}
\hline
Experiment & av. noise$^1$ & av. noise var.$^1$ & av. fgd res.$^1$ & $r_g$ &   $r$ \\\hline
A & 951.4 & 196.6 & 361.7 & 0.23 & 0.29\\ 
B & 18.6 & 15.3 & 13.8    & 0.03 & 0.05\\
C & 10.3 & 7.0 & 558.4    & 0.39 & 0.87\\
D & 13.0 & 4.3 & 3.2      & 0.0046 & 0.0056\\
\hline
A+WMAP8 & 942.2 & 190.7 & 212.1 & 0.15 & 0.18 \\
B+WMAP8 & 18.3 & 14.6 & 5.8     & 0.020 & 0.026\\
C+WMAP8 & 63.8 & 72.4 & 345.5   & 0.36 & 0.81\\
D+WMAP8 & 12.7 & 4.2 & 2.8      & 0.0044 & 0.0054\\
\hline
A+Planck & 389.2 & 79.0 & 43.2  & 0.07 & 0.10 \\
B+Planck & 18.3 & 14.7 & 8.3    & 0.024 & 0.035 \\
C+Planck & 23.2 & 16.0 & 2.8    & 0.018 & 0.019 \\
D+Planck & 12.8 & 4.2 & 3.0     & 0.0045 & 0.0055 \\
\hline
Planck & 1803.4 & 266.3 & 102.9 & 0.17 & 0.19 \\
\hline
\end{tabular}
\caption{
Notes --- $^1$ values in $10^{-4}\mu$K$^2$, see
  Section~\ref{sec:stat} for details.  
Here, $r_g$ \& $r$ are the tensor-to-scalar limits reachable by a given
experiment  assuming repectively that the residuals are an extra
Gaussian noise or a fixed foreground.
See Section~\ref{sec:stat} for details.
}
\label{tab:rmin}
\end{center}}
\end{table*}

\section{Experimental Optimization To Minimize Foregrounds}

In addition to studying the effect of foregrounds on each of these
four hypothetical experiments, we also designed an optimized
experiment to minimize the foregrounds. The optimization involves the
selection of frequency channels such that foregrounds are maximally
removed. Throughout this study we assume that the total number of
detectors in the focal plane is fixed to be 1000.

For this optimization, we extend the algorithm in
Ref.~\cite{Tegetal03} to allow for a non-uniform
weighting of the frequency bands. This is done by introducing a weight
vector ${\bf b}$ for the noise such that individual $b^i$ coefficients
are equal to the inverse of the square root of the number of detector
in each channel $i$. Equation~\ref{equ:alm} gets modified as,
\begin{equation}
\langle|a_{\ell m}|^2\rangle= {\bf w_\ell}^T{\cal S} {\bf w_\ell} + {\bf  w_\ell '}^T {\cal N} {\bf w_\ell '}  \, .
\end{equation}
Here, the matrix ${\cal S}$ is the signal (including foregrounds)
correlation matrix and ${\cal N}$ is the noise correlation matrix.   
The coefficients $w'^i_\ell$ are equal to $w^i_\ell\, b^i$. To
optimize the frequency channels, we fixed the normalization of the
$b^i$ coefficient so that $\sum_i{(b^i)^{-2}}=N_d$ where $N_d$
represents the total number of detectors and numerically solved for
the $b^i$ coefficient at each frequency. For each $b$ vector,
$w^i_\ell$'s are obtained by following the same procedure as before. 

We also include the sky coverage in our optimization study by
considering different options for sky coverage assuming total
integration time is fixed. In the context of B-mode observations, the
optimal sky area required to maximize the sensitivity to primordial
tensor modes has been studied in the literature.  With foregrounds
ignored and the optical depth to reionization assumed to be zero,
Ref.~\cite{KamJaf00} showed that a small patch of a few square
degrees may be adequate to detect primordial B-modes as the B-mode
spectrum peaks at $\ell$ of 100 associated with the projection of the
horizon size at the matter-radiation equality. With the bump at $\ell 
\sim 10$ included, for models with optical depth to reionization about
0.1 and higher, the optimal sky area becomes larger for CMB
polarization experiments that attempt to detect this bump and use it
as the primary means to detect tensor modes. Furthermore, if lensing
B-modes are the dominant contribution, then almost all-sky data are
required for an analysis of lensing confusion \cite{Kesden}.

However, in all these prior discussions related to sky coverage
optimization, contamination from foregrounds was ignored. 
We emphasize that with polarized foregrounds dominating the E-mode
primordial power spectrum at most frequencies and the B-mode power
spectrum for $r<0.1$ by a large factor at all frequencies, sky
coverage becomes inextricably linked to foreground removal. 
Instead of treating $f_{\rm sky}$ as a random variable, we consider
three specific options here by selecting 3 different sky areas for
optimization. These 3 sky coverages are simply areas of the sky where
the absolute value of the galactic latitude is higher than 20, 40, 60
degrees, and cover respectively 65.8, 35.7, 13.4\% of the entire
sky. Note that in each of the three cases, the total integration time
is fixed so that the noise in an individual pixel in smaller area maps
is lower relative to the pixel noise in a larger area map. We note
that the sky cuts we make are not necessarily optimal since the dust
and synchrotron emission are not symmetric with respect to the
galactic disk as can be seen in Figure \ref{fig:syncdustmaps}. We
leave this more detailed optimization to a future study.

\begin{table}
\begin{center}
\begin{tabular}{ccccccccc}
\hline
Freq. (GHz)& 30 & 45 & 70 & 100 & 150 & 220 & 340 & 500 \\
NET/det$^1$ & 71.7 & 60.1 & 50.7 & 46.0 & 46.0 & 60.1 & 172.5 & 1310.1\\
\hline \hline
$f_{\rm sky}$ & & & & & & & &  \\
\hline
65.8\% & 31 & 97 & 0  & 163 & 416 & 0 & 215 & 78 \\
35.7\% & 26 & 83 & 0 & 10 & 571 & 0 & 230 & 80 \\
13.4\% & 30 & 98 & 0 & 0 & 584 & 0 & 212 & 76 \\
\hline
\end{tabular}
\caption{Notes ---
Optimal detector distributions in our 8 frequencies for the different
sky coverages assuming the total number of detectors (1000) is
fixed. The distributions for the three sky coverages are similar and
such that  40-60\% of the detectors are assigned to CMB 150 GHz
channel, 30-35 \% at high frequencies to measure polarized dust, and
10-15 \% at low frequencies to measure the polarized synchrotron
radiation.  $^1$The tabulated NET values, per detector, are in
$\mu$K$\sqrt{\rm sec}$ and are consistent with values quoted for
bolometric observations from space \cite{Bock}.}
\label{tab:freqopti}
\end{center}
\end{table}

\begin{table*}
\begin{center}
\begin{tabular}{cccccccc}
\hline
& \bt Sky \\[-5pt] coverage\et & \bt average \\[-5pt] noise \et & \bt
average \\[-5pt] noise \\[-5pt] variance \et & \bt average \\[-5pt]
foreground \\[-5pt] residual \et & $r_g$ & $r$ & \bt optimization
\\[-5pt] improvement \et \\\hline  
Lensing & 65.8\%& 1000 & 150 & 180 & 0.0012 & 0.0014 & 1.72/1.02\\
Ignored &35.7\% & 483  & 100 & 131 & 0.0011 & 0.0014 & 1.69/1.27\\
& 13.4\% & 172 &  58 &  41   & 0.0007      &  0.0015 & 1.85/1.78\\\hline\hline
With & 65.8\%& 1000 & 150 & 180 & 0.0027 & 0.0018 & 1.72/1.27\\
Lensing & 35.7\% & 483  & 100 & 131 & 0.0036  &  0.0028     & 1.69/1.23\\
& 13.4\% & 172 &  58 &  41   & 0.0054      &  0.0132 & 1.85/1.09\\\hline
\end{tabular}
\caption{Average (between $\ell = 2$ and $100$) level of residual
foreground and noise ($C_\ell(\ell+1)\ell/2\pi$, units are in nK$^2$)
after foreground removal for different sky coverages. The ``average
noise variance'' represents the noise residual once an estimate of the
noise power spectrum is subtracted.  $r_g$ \& $r$ are the
tensor-to-scalar limits reachable by a given experiment assuming
repectively that the residuals are an extra Gaussian noise or a fixed
foreground. We quote $r_g$ for comparison but stress that it is
unlikely that the residuals can be treated as extra Gaussian noise.
See Section~\ref{sec:stat} for details. The optimization improvement
values are the ratio of foreground residual (left) and of $r$ (right)
between the non-optimized and optimized setup. The top three lines
ignore the additional noise related to lensing B-modes, while the
bottom three lines calculate the minimum tensor-to-scalar ratios with
lensing B-modes included as a residual noise so that
$R_\ell=R_\ell^{\rm fore}+C_\ell^{\rm BB,lens}$.  As tabulated, B-mode
lensing confusion leads to a factor of 2 to 8 degradation in the
minimum tensor-to-scalar ratio suggesting that ``lens cleaning''
techniques can be implemented to recover this reduction in part.}
\label{tab:opti}
\end{center}
\end{table*}

\subsection{Optimization Results}

Following the above procedure, we optimized the channel selection for
the three sky cuts mentioned above and by making use
of the single between $\ell$ of 2 and 100 to detect primordial B-modes.
The ``optimal'' frequency
distribution (see, Table~\ref{tab:freqopti}) is characterized by the
dominance of the ``CMB channel'' (here 150 GHz), which represents 40
to 60\% of the total number of detectors, with adequate coverage at
both high and low frequencies for removal of dust and synchrotron
respectively. Note that this optimization does not include data from
other experiments since we have already seen that experiment D, which
is our example space-based experiment, shows no improvement when
combined with either 8 years of WMAP data or Planck data.  As
tabulated in Table~\ref{tab:freqopti}, frequency optimization is such
that the number of detectors is higher in the primary CMB channel of
150 GHz than the number of detectors in other channels. In terms of
the overall focal plane sensitivity, the fractional contribution from
detectors at 150 GHz is even higher since the instrumental noise for
detectors at this frequency is the lowest (see second line of
Table~\ref{tab:freqopti}).

As shown in Table~\ref{tab:freqopti}, our algorithm also selected the
two channels at both the lowest and highest frequencies to remove
synchrotron and dust, representing respectively about 10 to 13 \% and
28 to 32 \% of the total number of detectors. This arrangement
is necessary to study how the spectrum changes across the sky and the
number of detectors is selected in such a way that the resulting noise
associated with the uncertainty of the spectral indices of either one
of the two foregrounds does not dominate the overall noise. We remind
the reader that our spectral indices vary across the sky but they
don't change with frequency.  In these foreground channel pairs, the
dominant ones are 45 GHz and 340 GHz (in terms of the number of
detectors at those channels) because of their raw sensitivity, and the
large number of detectors compensates for a smaller arm leverage (see
Figure \ref{fig:spectra}).  Furthermore, in our optimization
almost 50\% of the detectors go to the channel at 150 GHz which acts
as the primary channel for CMB measurements. 

The achievable limits on the tensor-to-scalar ratio are highlighted in
Table~\ref{tab:opti}. We tabulated results with and without lensing
B-modes as a source of noise. If lensing B-modes could be reduced to a
level lower than the residual foreground, the minimum $r$ achievable
is 1.4$\times 10^{-3}$ obtained by covering 36\% or 66\% of the sky.
Including lensing as an irreducible noise such that
$R_\ell=R_\ell^{\rm fore}+C_\ell^{\rm BB,lens}$ leads to minimum
tensor-to-scalar ratios of 1.8$\times 10^{-3}$, 2.8$\times 10^{-3}$,
and 1.3$\times 10^{-2}$ for respectively 65.8, 35.7, 13.4 \% sky
coverage. The 66\% sky coverage is more efficient on these very large
scales, but the lensing B-modes provide a floor to the primordial
B-mode detection there. However on these very large scales, this
limit is sensitive to the exact shape of the B-mode reionization bump
which in turn depends on the optical depth and the reionization
history.  In general, our limits show that the suggested $r < 0.01$
goal of the next generation {\it Inflation Probe} mission (see the
report of the Task Force on CMB Research \cite{Weiss}) is achievable.

In the last column of Table~\ref{tab:opti}, we highlighted the
improvement on the foreground residual and the minimum $r$ when the
experiment is optimized relative to an experiment where all the
detectors are equally distributed in number between the 8 channels,
with 125 detectors per channel.  As tabulated, optimization leads to
roughly a 70\% improvement in the foreground residual and a 2 to 80\%
improvement in the minimum tensor-to-scalar ratio measurable with each
of the three options without lensing contamination. However when the
lensing is the limiting factor, the 70\% gain on the foreground level
only leads to a 9\% to a 27\% improvement.

As discussed above, in addition to residual foreground noise, with a
high sensitivity space-based experiment one must also account for the
lensing B-mode signal that acts as another source of confusion
\cite{Kesden,Song,Seljak:2003pn}. In Table~IV, we also listed the
limits on tensor-to-scalar ratio when lensing is ignored so that one
can compare the degradation related to lensing alone.  As shown by the
difference, with lensing included as a source of noise, the resulting
constraints on the tensor-to-scalar ratio is a factor of 3 to 6 worse
compared to the case when the confusion from lensing B-modes is not
present.  We highlighted this primarily due to the fact that
techniques have been developed to reconstruct the lensing deflection
angle and to partly reduce the B-mode lensing confusion when searching
for primordial gravitational waves
\cite{HuOka02,Kesden03,Hirata03}. These construction techniques
make use of the non-Gaussian signal generated in the CMB sky by
gravitational lensing.  The lensing signal has a unique
non-Gaussian pattern captured by a zero three-point correlation
function and a non-zero four-point correlation function. This
non-Gaussian feature will be very useful for cleaning the lensing 
signal but detrimental when trying to measure it because of the
increased sample variance \cite{Smith04,Smith06}. 
However, residual foregrounds are also non-Gaussian and they could
bias the extraction of the lensing signal. Thus, in addition to an
increase in the power-spectrum  noise, foregrounds may also impact
proposed techniques to clean the lensing signal. In an upcoming paper,
we will discuss the impact of residual foregrounds on these lensing
analyses.  

\section{Conclusions}

The B-modes of Cosmic Microwave Background (CMB) polarization may
contain a distinct signature of the primordial gravitational wave
background that was generated during inflation and the amplitude of
the background captures the energy scale of inflation.
Unfortunately, the detection of primordial CMB B-mode polarization is 
significantly confused by the polarized emission from foregrounds,
mainly dust and synchrotron within the galaxy. Based on
polarized maps from WMAP third-year analysis, which include
information on synchrotron radiation at 23 GHz, and a model of 
the polarized dust emission map, we considered the ability of
hypothetical CMB polarization experiments with frequency channels
between 20 GHz and 300 GHz to separate primordial CMB from galactic
foregrounds. Planck data will aid experiments with narrow frequency
coverage to reduce their foreground contamination and we found 
improvements of factors of 2 to 40 in their achievable $r$.

We also studied an optimization of the distribution of detectors among
frequency channels of a CMB experiment with a fixed number of
detectors in the focal plane. The optimal configuration (to minimize
the amplitude of detectable primordial B-modes) requires observations
in at least 5 channels widely spread over the frequency range between
30 GHz and 500 GHz with substantial coverage at frequencies around 150
GHz. If a low-resolution space experiment concentrates on roughly 66\%
of the sky with the least contamination, and with 1000 detectors reach
a noise level of about 1000 (nK)$^2$, the minimum detectable level of
the tensor-to-scalar ratio would be about 0.002 at the 99\% confidence
level. The bulk of the sensitivity comes from the largest angular
scales, i.e., using the signal created during reionization.

Our results indicate that the goal for the next generation {\it
Inflation Probe} mission as outlined in the recent report from the
Task Force on CMB Research \cite{Weiss} is achievable. 

{\bf Acknowledgments:} We would like to acknowledge the use of the
HEALPix package for our map pixelization \cite{Gorski} and
participants of the Irvine Conference on ``Fundamental Physics with
CMB Radiation'' for useful discussions and suggestions that led to
this work.  We thank the WMAP flight team and NASA office of Space
Sciences for making available processed WMAP data public from the
Legacy Archive for Microwave Background Data Analysis
(LAMBDA)\footnote{http://lambda.gsfc.nasa.gov}. This work is
supported by a McCue Fellowship to AA at UC Irvine Center for
Cosmology.

\end{document}